\documentclass[aps,twocolumn,floatfix,nofootinbib]{revtex4}

\usepackage[T1]{fontenc}
\usepackage{color}
\usepackage[latin9]{inputenc}

\usepackage{textcomp}
\usepackage{appendix}
\usepackage{amstext}
\usepackage{amssymb}
\usepackage{graphicx}
\usepackage{setspace}

\usepackage{graphicx,bm,psfrag, amsmath, bbm}

\newcommand{\beq}{\begin{equation}}
\newcommand{\eeq}{\end{equation}}
\newcommand{\beqa}{\begin{eqnarray}}
\newcommand{\eeqa}{\end{eqnarray}}
\newcommand{\ba}{\begin{array}}
\newcommand{\ea}{\end{array}}

\usepackage{youngtab}

\newcommand{\be}{\begin{equation}}
\newcommand{\ee}{\end{equation}}
\newcommand{\bea}{\begin{eqnarray}}
\newcommand{\eea}{\end{eqnarray}}
\newcommand{\Tr}{{\rm{Tr}}}

\begin{document}

\title{\boldmath Novel Higgs Potentials from Gauge Mediation of Exact Scale Breaking}

\author{ Steven Abel and Alberto Mariotti}
\affiliation{Institute for Particle Physics Phenomenology,
Durham University, South Road, Durham, DH1 3LE}

\date{\today}

\begin{abstract}

\noindent 
We present a gauge mediation principle for BSM theories where {\em exact} UV scale invariance is broken in a hidden sector. The relevant configurations are those in which the Standard Model and a hidden sector emanate from a scale invariant pair of UV theories that communicate only via gauge interactions.
We compute the radiatively induced Higgs potential which contains logarithmic mass-squared terms that lead to unusual Higgs self-couplings. Its other couplings are unchanged.
\end{abstract}

\maketitle


\section{Introduction and overview}

The existence in Nature of mass scales that are much smaller
than the Planck scale suggests an additional symmetry at work in the
Standard Model. Lately the old idea that the relative lightness of
the Higgs has something to do with scale-invariance \cite{Salam,Coleman:1973jx,gildener,bardeen}
has been gaining some ground \cite{scaleinv1,scaleinv2,scaleinv3,Goldberger:2007zk,scaleinv4,scaleinv5,scaleinv6,scaleinv7,scaleinv8,scaleinv9,scaleinv10,scaleinv11,scaleinv115,scaleinv12,scaleinv13,scaleinv14,scaleinv15,scaleinv17,scaleinv18,scaleinv20,scaleinv21,scaleinv22,late1,late2,late3,late4,late5,late7,late9,late10,late11,late12,late13,late14,late15,late16,late17,late18}.
 
It has to be said that scale invariance is not an obvious candidate for a symmetry to protect the Higgs, 
because the Standard Model has scale anomalies. In the absence of
a Higgs mass the theory is only ``classically scale invariant''
 (which is to say that it isn't), and {anomalous symmetries only buy a loop of protection}. 
Of course scaling anomalies are precisely the starting point of the Coleman-Weinberg
mechanism \cite{Coleman:1973jx,gildener}, but whether or not classical scale invariance
can have a well-defined meaning in a UV complete theory is still a matter for debate \footnote{More precisely it is difficult to see what symmetry in the UV complete theory could guarantee a low energy theory with classical scale
invariance, when exact scale invariance does not.}. 

 The notion of scale invariance is clearly on firmer footing if the Standard Model (SM) ultimately emanates from a theory
that is \emph{exactly} scale invariant in the ultraviolet UV. 
This is a fixed-point of the renormalisation group (RG) which means that the theory stops running at very high energies.
The couplings and anomalous dimensions will change as the theory flows towards the infra-red (IR), but 
the fact that it flows out of a UV fixed point renders all UV divergences harmless. 
This idea was pioneered in the context of asymptotic safety \cite{asymp-safety} (for a recent review see \cite{Litim:2011cp}),
and may have relevance in other scenarios including a number that address the hierarchy problem (for example 
\cite{Gies:2003dp, Gies:2009sv, Braun:2010tt, Gies:2013pma, Bazzocchi:2011vr}). General aspects of this idea were 
recently discussed in \cite{late7}.

Unfortunately finding calculable predictions in such theories is generally difficult because it is not possible to follow the renormalisation group 
flow analytically from the UV fixed point all the way to the IR. However here we will present one situation that {\em can} give perturbatively calculable results, and 
we will calculate the resulting Higgs potential which deviates substantially from the SM one. We suppose firstly that at the UV fixed point there are additional degrees of freedom that communicate to the SM only through its gauge couplings. Secondly, we assume that they dominate the RG flow away from the fixed point, while the SM components and gauge couplings remain almost stationary. 

This type of situation is referred to as a ``saddle-point'', and the previous statement amounts to saying that the UV fixed point  is ``attractive'' in the SM directions and 
``repulsive'' in the directions of the additional degrees of freedom. As we shall see, such systems can easily be  constructed by sewing together two theories, an SM augmented so that it  
has an attractive (IR-stable) fixed point, and another theory that has a repulsive (UV-stable) fixed point. The latter we will refer to as the ``hidden sector''.

One expects that the dominant RG flow in the hidden sector can lead to spontaneous breaking of scale invariance. 
In other words it causes some scalar to acquire a VEV, $f_c$. 
This breaking may arise from radiative terms, 
or it may arise in the context of some scale invariant UV complete theory, for example NJL models of strongly coupled fermionic systems \cite{Gies:2003dp,Barnard:2013zea}.
As there are only gauge couplings between these degrees of freedom and the SM, the spontaneous breaking of scale breaking leaks through to the SM in a controlled way. In particular it yields calculable and novel results for the Higgs potential. 
 
\begin{figure}
\noindent 
\begin{center}
\includegraphics[scale=0.4]
{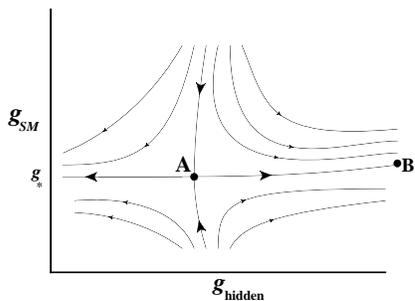}~~~~~~~~~~~~~~~~~~~\caption{Schematic depiction of the UV fixed point structure required for GMESB. The SM couplings  $g_{SM}$
and anomalous dimension are attracted to the fixed point at $A$, while the additional (hidden) sector to which it couples only via gauge interactions initiates a flow 
to point $B$ where it spontaneously breaks scale invariance.
 \label{fig:saddle} }
\end{center}
\end{figure}

 This configuration, depicted in figure~\ref{fig:saddle}, is tractable because it is modular, in the sense that gauge mediation of supersymmetry breaking (GMSB) is modular (for a review see \cite{Giudice:1998bp}), and therefore we refer to the framework as Gauge Mediated Exact Scale Breaking (GMESB). Its principle can be stated as follows:\\
\\
\emph{The SM resides in an extended conformal theory in the UV. It couples only via gauge interactions to a sector that 
initiates flow away from the fixed point and spontaneously breaks scale invariance at a scale $f_c$.} \\

\noindent 

Readers familiar with spontaneously broken  scale invariance will recognise the field $\chi$ getting the VEV $\langle\chi\rangle = f_c$ as its 
Goldstone mode, namely the dilaton. Ultimately, since the Higgs itself gets a (much smaller) VEV, the
\emph{actual} dilaton would be composed mainly of $\chi$ with a small
admixture of Higgs. Therefore the problems that plague the usual Higgs-as-dilaton
idea \cite{Goldberger:2007zk}
do not apply. In particular it would have all the same couplings
to the other particles of the SM -- although as we shall see its self-couplings
are generally different.
\begin{figure}
\noindent \centering{}~\includegraphics[scale=0.3]
{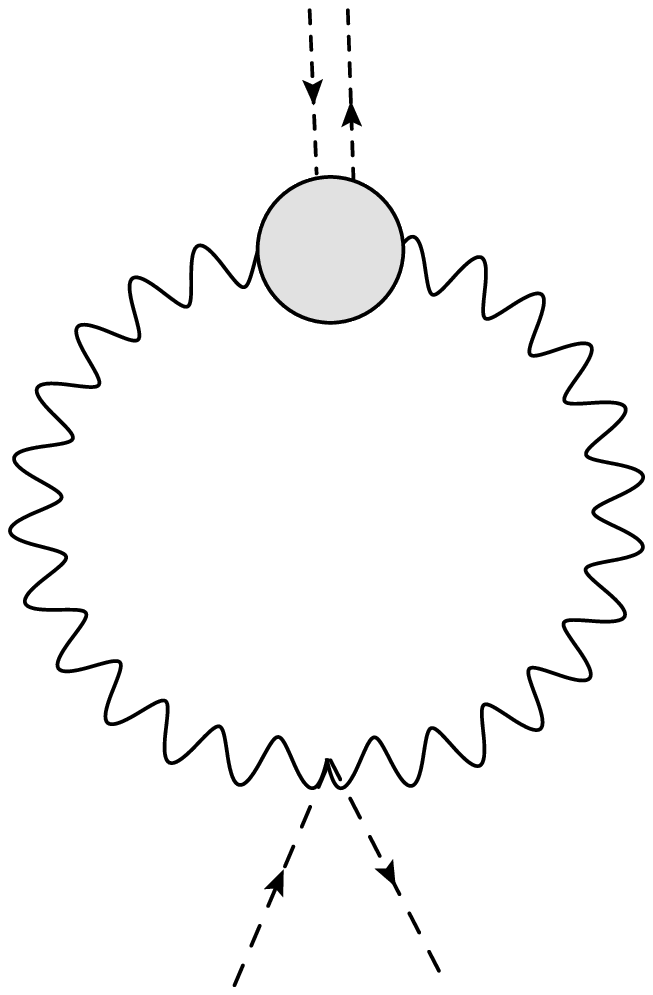}~~~~\includegraphics[scale=0.32]{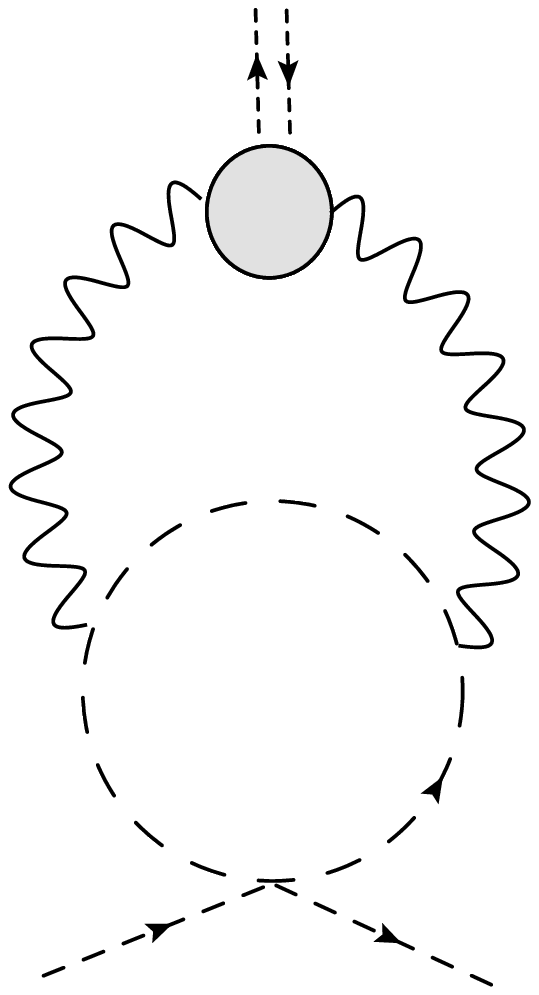}~~\includegraphics[scale=0.32]{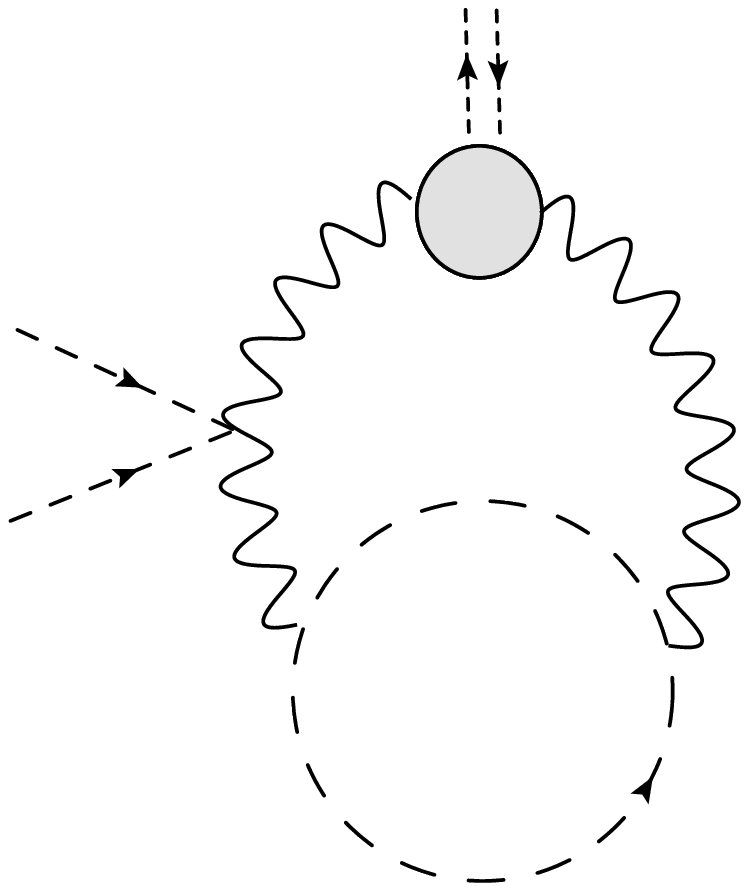}~~~~\includegraphics[scale=0.35]{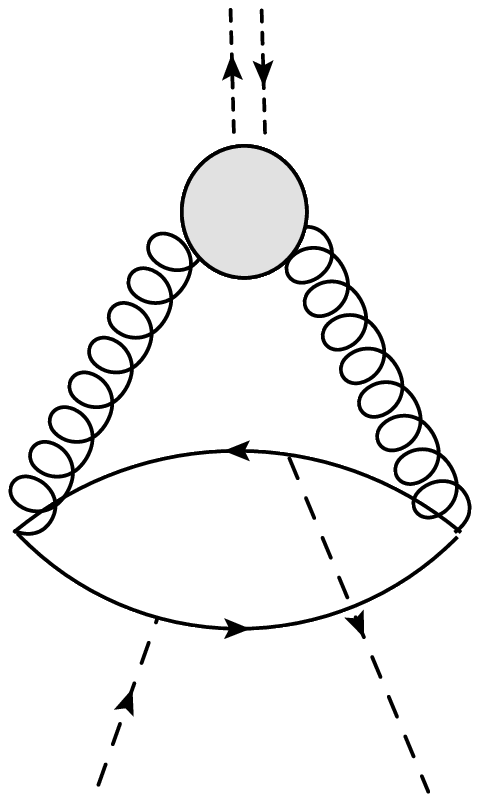}\caption{Leading radiative contributions to the Higgs mass squared coupling. The blob represents the scale anomaly
of the $SU(2)_{W}$ and $SU(3)_{c}$ gauge groups respectively.\label{fig:Contributions-to-the} }
\end{figure}

The obvious example of such a system would be if the SM gauge couplings ran
to zero at a Gaussian UV fixed point. In this case the Yukawa couplings and
Higgs self interactions would also run to zero, and the SM would just
become a free field theory in the UV. In this case, UV divergent contributions
to the Higgs mass would be tamed by the mediation itself flowing to 
zero in the UV. (This is in essence the approach discussed in \cite{Wetterich:2011aa} and evoked in the 
discussion of \cite{scaleinv13}.) This case would be difficult to treat analytically. 

The case of interest here is rather easier to analyse: it has a UV fixed point that is
interacting, with non-zero couplings and anomalous dimensions.
When $\langle\chi\rangle\rightarrow0$, scale invariance
is broken only by the Higgs VEV which in that limit becomes the true dilaton of the theory.
All the couplings in the SM remain at their UV fixed point values
$g_{*}$ and therefore the Higgs mass itself is zero to all orders.
Once $\langle\chi\rangle=f_{c}$
is turned on and becomes dominant (so that the Higgs is no longer the true dilaton) we shall find, 
by expanding about $\langle\chi\rangle=0$,
 that the effective
low energy theory has relevant operators proportional to $f_{c}$.
(The leading contributions to the Higgs mass come from would-be logarithmic 
UV divergences that are tamed by small non-zero anomalous dimensions at the UV fixed point.)

One might suppose that the phenomenology of such a theory would be
Coleman-Weinberg like in the IR, but it is not. In accord with our
proverb about anomalous symmetries,
one finds all kinds of classically dimensionful operators in the IR, of
order $f_{c}$ suppressed by loops. This illustrates the general fact that
classically scale invariant theories and truly scale invariant theories
are not generally speaking close cousins. Crucially, if there is ever a return to exact scale invariance 
in the UV, the scale at which that happens will be the one that governs the relevant operators in the effective IR theory.

In particular we will show that 
the effective IR theory has a potential containing a logarithmic Higgs mass-squared term:
\bea
V &=&
\frac{\lambda}{4} \phi^4+
\frac{1}{4} \phi^2  \left(-m_h^2 +\left(m_h^2-2 \langle \phi  \rangle^2 \lambda \right) \log \left[\frac{\phi^2}{\langle \phi \rangle^2}\right]\right), \nonumber \\
&& m_{h}^{2} \sim  \mbox{Loop-factors}\times f_{c}^{2}.\label{eq:rad0}
\eea
Note that there will be the usual logarithmic terms for the quartic coupling as well, but their effect on electroweak symmetry breaking is negligible. The coupling $\lambda$ is arbitrary as it can be present at the UV fixed point, whereas we will derive $m^2_h$ and $\langle\phi\rangle$ from radiative corrections. 

Expanding the potential (\ref{eq:rad0}) about the minimum as $\phi = \langle \phi \rangle +h$, 
the Higgs self-couplings are given by 
\bea
\label{Vhiggs}
V \hspace{-0.1cm} &=&
\hspace{-0.1cm} \frac{1}{4} \left(\langle \phi \rangle^4 \lambda -m_h^2 \langle \phi \rangle^2\right)+\frac{m_h^2 h^2}{2}+\left(\frac{m_h^2}{6 \langle \phi \rangle}+\frac{2 \langle \phi \rangle \lambda }{3}\right) h^3 \nonumber \\
&& \hspace{1cm}+\left(\frac{\lambda }{3}-\frac{m_h^2}{24 \langle \phi \rangle^2}\right) h^4 + {\mathcal O}(h^5)\, .
\eea
The SM potential and couplings are recovered in the limit that one chooses the SM value of $\lambda = \frac{m_h^2}{ 2\langle \phi \rangle^2}$. An interesting alternative however is 
$\lambda = \frac{m_h^2}{\langle \phi \rangle^2}$, which gives an electroweak breaking minimum that is degenerate with the symmetric minimum at the origin. One can of course arrange a long lived metastable minimum as well. Finally, if $\lambda$ is negligible, the potential takes the ``running-Higgs-mass-squared'' form, comparable to that of the simplest (e.g. $UDD$) $F$- and $D$-flat directions in the MSSM: 
\begin{equation}
V=\frac{1}{4}m_{h}^{2}\phi^{2}\left(\log\frac{\phi^{2}}{\left\langle \phi\right\rangle ^{2}}-1\right) \, . \label{eq:rad1}
\end{equation}

Roughly speaking the appearance of the logarithmic potential can be
understood as follows. Any field that changes the $\beta$ function
couples through the scale anomaly to the corresponding gauge bosons.
Therefore the Yang-Mills terms in the effective Lagrangian pick up
contributions of the form 
\begin{equation}
\mathcal{L}\supset\beta(\phi,\chi)FF,
\end{equation}
where the $\phi$ dependence is dominated by the familiar top quark
and Higgs contributions to the beta functions. The $\chi$ dependence
arises from the running to the fixed point around the scale $\langle\chi\rangle=f_{c}$.
In some scenarios $\chi$ could be a non-dynamical auxilliary field
(i.e. a spurion) encoding for example fermion masses. It will not
be crucial for our discussion if $\chi$ is a dynamical field or
not. 

At leading order the contributions to the gauge
propagator can be split into the visible and hidden parts \footnote{We should add 
a word about nomenclature; here and throughout, by ``hidden'' we will mean fields that couple only to the gauge bosons of the Standard Model, or not at all. When we refer specifically to the former we will call them messengers, in the tradition of GMSB.}, ${C}_{}\supset{C}_{vis}(\phi)+{C}_{hid}(\chi)$.
One should recall that in non-supersymmetric systems the one-loop gauge propagator is not proportional to 
the beta function (as the latter also gets vertex contributions) so the analysis is unfortunately more complicated than simply putting the 
dressed propagators into the perturbative expansion. Nevertheless, higher order terms begin
with the cross-term ${C}_{}\supset{C}_{vis} {C}_{hid}$.
Diagrams that do not involve ${C}_{}$ at all have no dependence
on $\phi$ or $f_{c}$ and therefore cancel at the fixed point \footnote{Note that the couplings should all take their UV fixed point values.
Any deviation would also be a function of $\phi$ and $f_{c}$ so
would be automatically included in this perturbative expansion.%
}. The vanishing of the beta functions at the UV fixed point guarantees that terms 
in the effective potential derived from 
${C}_{vis}(p)$
can contribute only to finite quartic terms for
the visible sector Higgs. Finally the interesting new
terms arise from the cross-terms (effectively the 2nd, 3rd and 4th diagrams of 
figure \ref{fig:Contributions-to-the}), and from
2-loop diagrams like the first diagram of fig.\ref{fig:Contributions-to-the}. 
It turns out that the two-loop and three-loop-top diagram have no
logarithmic terms; they are effectively computable portal terms. 
The logarithmic piece in the
potential derives from two diagrams involving the Higgs loop.
The coefficient of all these terms is relatively
small so that the potential
naturally has a minimum at $\langle\phi\rangle\ll f_{c}$.
Note that, although the minimum is generated radiatively, the obvious advantage over the visible-sector Coleman-Weinberg mechanism is that
$\langle\phi\rangle$ and $f_{c}$ are essentially independent
parameters, and so one avoids the problematic loop-suppressed Higgs mass. 

In the next section we first derive the effective potential by carefully analysing
the contributions from the diagrams of figure 1.
In the section that follows we discuss the global configuration, presenting 
a simple perturbative example of a QCD theory with a UV
fixed point. Finally we discuss the phenomenological aspects, including the expected scale of new physics, which turns
out to be naturally of order $1$--$10^3$ TeV.

\section{The Calculation}

\label{GMESB}

As described in the Introduction, the set-up we consider is a theory that can be split into a visible (SM) sector and a hidden sector, by sending the gauge coupling to zero. The hidden sector consists of a 
part that is responsible for spontaneously breaking scale invariance at a scale $f_c$, 
and a messenger sector containing fields that are charged under the SM gauge groups, to which it directly couples. 
The gauge group of the SM appears as a global symmetry to the hidden sector dynamics that causes breaking of scale invariance. 
At a conformal fixed point (where $f_c=0$) any non-gauge (e.g. portal-like) coupling connecting the visible and hidden sectors is vanishing.

As we mentioned, such a configuration can be built by considering the properties of the visible+messenger theory (i.e. turn off the hidden sector gauge groups) and 
hidden+messenger (i.e. turn off the SM gauge groups) independently. The latter should have a UV stable fixed point while the visible+messenger sector should be a scale invariant version of the SM, which possesses a non trivial IR stable fixed point. (Thus when all gauge couplings are turned on one expects to find the saddle point behaviour described above with the hidden sector initiating the flow away from the fixed point in the IR). Any portal-like coupling connecting the visible and hidden sectors are vanishing at the UV fixed point \footnote{There are two possible reasons for 
the portal coupling to vanish there. The first is simply that one solution to all beta functions vanishing has zero portal coupling. The second would be a symmetry 
argument. For this one might consider the fact that in the limit $f_c \to 0$ and $g_{SM} \to 0$ there is an effective enhancement
of conformal symmetry to two conformal symmetries in the two sectors, much in the spirit of
\cite{Benakli:2007zza,Cheung:2010mc,Argurio:2011hs,late14}.}.

In the simplest case, the SM can be made conformal by adding extra states that 
couple only via gauge interactions to the SM (and not via Yukawa couplings). 
If these states couple to the hidden sector as well then they act as 
messengers. This is of course trivial to 
arrange (with direct couplings to the SM forbidden with symmetries) for the non-abelian factors of the SM gauge group. For example vector-like pairs of $SU(3)$ triplets and $SU(2)$ doublets suffice. 
Doing the same for the hypercharge coupling would require some kind of unification, so we shall not attempt to build a complete model here. 

 
Despite that, we shall in the 
next section present a toy configuration of an $SU(3)$ model with an IR fixed point coupled via bi-fundamental fermions (messengers) to a ``hidden'' $SU(N)$ model with a repulsive UV fixed point, that has all the 
desired properties. In particular this minimal example shows that one does not necessarily expect to find additional states at mass scales much below $f_c$. 

On the other hand, more elaborate implementations could include additional states in the visible sector to make it conformal, for example  coloured scalars coupling to the Higgs through quartic interactions.
Those states should not acquire a mass of order $f_c$ (i.e. they should not couple to the hidden sector directly as well), otherwise scale breaking would be 
directly mediated to the Higgs. They would be expected to acquire masses proportional to  $f_c$ but loop suppressed, much like the Higgs itself. This is of course also analogous to GMSB, the new extra states being similar to gauginos and sfermions which acquire 
masses comparable to the Higgs through gauge mediation.
Here we shall focus on the simplest case and assume such states are absent.

Let us now proceed with the computation
of the Higgs potential by the spontaneous breaking of scale invariance in the
hidden sector.  We allow the Higgs to have a tree level quartic potential with coupling $\lambda$,
consistent with the requirements of conformal symmetry. As alluded to in the Introduction our goal is to obtain calculable expressions, so we assume that the SM gauge couplings are always perturbative.  This makes it possible to evaluate the effective potential for the Higgs by considering loops 
of SM gauge fields, with insertions of current correlators. From now on we neglect the contributions mediated by $U(1)_Y$.

 The Standard Model gauge bosons couple perturbatively both to the currents of the Standard Model matter
$J^{\mu}_{vis}$ and to the currents representing the hidden sector $J^{\mu}_{hid}$ as
\be
\mathcal{L} \supset g  A_{\mu}^a ( J^{\mu a}_{hid}+J^{\mu a}_{vis})
\ee
where here $A_{\mu}^a$ can represent the $W$ bosons or the gluons for the $SU(2)$ and $SU(3)$ case respectively,
with $a$ running over the generators of the gauge group.
Henceforth we work with the generic case of an $SU(N)$ gauge group. We will specialise to $SU(2)$ and $SU(3)$
at the end of the computation.

The fact that the two currents are not directly coupled but communicate only through gauge interactions can be taken as the formal definition of gauge mediation.
We parameterise the two point functions of the currents with dimensionless functions\footnote{We work in Euclidean space and unitary gauge.}, $C$;
\bea
&&
\langle J^{\mu a}_{vis} J^{\nu b}_{vis} \rangle=-(p^2 \eta_{\mu \nu}-p_{\mu} p_{\nu}) C_{vis}^{ab}(p^2, \phi^2)\\
&&
\langle J_{hid}^{\mu a} J_{hid}^{\nu b} \rangle=-(p^2 \eta_{\mu \nu}-p_{\mu} p_{\nu}) C_{hid}^{ab}(p^2,f_c^2)\, .
\eea
The visible sector two point function depends on the Higgs field vev,
here denoted $\phi$,
while the hidden sector one depends on the characteristic scale $f_c$.
For simplicity we take the two point functions to be diagonal in the non-abelian indices.
We also assume that these functions depend only quadratically on $\phi $ and $f_c$.

In this parameterisation, the current two point functions correct the
propagator of the gauge boson producing an effective potential induced by a loop of gauge boson;
\be
\label{Veff0}
V_{eff}=\frac{3}{2} \Tr \int \frac{d^4 p}{ (2 \pi)^4}  \log \left(1+ \frac{m_V^2}{p^2}+ g^2 C_{vis} +g^2 C_{hid}\right),
\ee
where we denoted by $m_V$ the gauge boson mass,
relevant for the electroweak case.
The trace is over the gauge group indices; 
for $SU(N)$ gauge groups and diagonal $C$ functions and mass this gives an overall $N^2-1$ factor.

In evaluating the possible quantum corrections, we will of course be focussing on the terms that are proportional to
$f_c^2 \phi^2$, which will determine the quadratic part of the potential for the Higgs.
There are other  contributions, resulting for example from the discontinuities in anomalous
dimension induced by the Higgs vev, which can generate only quartic terms for the Higgs. 
These terms are negligible in determining the minimum.

Our gauge mediation principle implies that in the limit $f_c \to 0$ the effective potential mixing the
two sectors should vanish.
We can enforce this, and at the same time select only the contributions of interest,
by computing the difference between the effective potential (\ref{Veff0}) with $f_c \neq 0$ and the effective potential with $f_c=0$;
\be
\label{pipo}
\delta_{f_c} V_{eff}=V_{eff} (f_c^2)-V_{eff}(f_c^2=0)\, .
\ee
Note that in each term there are divergences that do not
involve the scale $f_c$. These correspond to wave function renormalisation and simply set the anomalous dimensions at the conformal fixed point and cancel in eq.(\ref{pipo}). Of course the advantage of the GMESB approach is that we do not need to consider them.

Expanding the effective potential in the gauge coupling
we obtain at ${\cal O}(g^4)$
\bea
\delta_{f_c} V_{eff} \hspace{-0.1mm} &=& \hspace{-0.1mm} \frac{3 (N^2-1)}{2} \int \frac{d^4 p}{ (2 \pi)^4}  
\frac{p^2}{p^2+m_V^2} (g^2 \delta_{f_c} C_{hid} \nonumber \\
&& -
\frac{g^4}{2} (2 C_{vis}  \delta_{f_c} C_{hid}+  ( \delta_{f_c} C_{hid})^2 )
)\, .
\eea
The quadratic terms for the Higgs field $\phi$ appear here in the massive propagator for the gauge boson or inside the
$C_{vis}$ function.
Of the former contributions we need keep only the leading order ones, i.e. those 
at ${\cal O}(g^2)$ inside the parenthesis. We arrive at the final generic expression 
for the Higgs potential induced by an $SU(N)$ gauge group;
\bea
\label{DVeff}
\delta_{f_c} V_{eff} &=&\frac{3 (N^2-1)}{2} \int \frac{d^4 p}{ (2 \pi)^4}  
\left(
g^2 \frac{p^2}{p^2+m_V^2} \delta_{f_c} C_{hid}\right. 
\nonumber \\ 
&& \hspace{1cm}-
 g^4  C_{vis}(\phi^2)  \delta_{f_c} C_{hid}
\mbox{\LARGE )}\, .
\eea
Note that the first term in this expression gives rise to a Higgs mass proportional to $f_c$ once we expand the propagator. 
As we shall now see, the second term can give rise to both mass terms and logarithms, due to the dependence
of $C_{vis}$ on the Higgs field.

The form of the function $C_{vis}(p^2,\phi^2)$ is completely determined by the visible sector and changes if there are extra states
(massless at tree level) that couple to the Higgs. As mentioned above, in this paper we will explicitly consider only the simplest case where such states are absent. The function $C_{hid}$ is an unknown function provided by the hidden sector.
The hidden sectors can be different for the $SU(2)$ or $SU(3)$ gauge groups, having
 in principle different matter content and also dynamical scales.
We distinguish between them with an extra index, e.g. $\delta_{f_{c(2)}} C_{hid}^{(2)}$
for the hidden sector of $SU(2)$.

The main contribution to the integral of (\ref{DVeff}) comes from the 
 large momentum region, with $p^2 \gg \phi^2$.
Hence, in order to find an approximate expression for $\delta_{f_c} V_{eff}$
that already encodes all the relevant features, we 
can use the one loop large momentum expansion of the $C_{vis}$, namely
\cite{PassarinoBook}
\bea
&&
8 \pi^2 C^{(2)}=\log(\frac{\mu^2}{p^2}) b'^{(2)}+ \frac{m_H^2}{ 4 p^2}(1+\log \frac{m_H^2}{p^2})
\nonumber \\
&&\hspace{1cm} -6 \frac{m_t^2}{p^2} +\frac{m_W^2}{4 p^2}(51-13 \log \frac{m_W^2}{p^2})+{\mathcal O}(1/p^4) \nonumber \\
&&
\label{Cvislarge}
8 \pi^2 C^{(3)}= \log(\frac{\mu^2}{p^2}) b'^{(3)}-6 \frac{m_t^2}{p^2} +{\mathcal O}(1/p^4)
\eea
for $SU(2)$ and $SU(3)$ gauge bosons respectively.
Here $b'^{(2)}$ and $b'^{(3)}$ are the wave function renormalisation coefficients of the $SU(2)$ 
and $SU(3)$ gauge bosons.
The dependence on the Higgs doublet is encoded in the masses as 
$m_H^2= 4 \lambda H H^{\dagger} $, $m_W^2=\frac{g_2^2}{2} H H^{\dagger}$ and $m_t^2=\lambda_t^2 H H^{\dagger}$.

Inserting these expression into (\ref{DVeff}), expanding the massive propagator, and keeping only the
terms quadratic in the Higgs field, we get for the $SU(2)$ induced effective potential
\begin{widetext}
\be
\delta_{f_c} V_{eff}^{(2)}= 
\frac{9 g_2^4 \mathcal{A}_2 }{16 \pi^2}   f_{c(2)}^2
H H^{\dagger}
\left( 
4 \pi^2  -6 \lambda_t^2+\lambda (1+\frac{\mathcal{B}_2}{\mathcal{A}_2})
+\frac{51}{8} g_2^2 (1-\frac{13}{51}\frac{\mathcal{B}_2}{\mathcal{A}_2})
\label{effV2}
+\lambda \log \frac{4 \lambda H H^{\dagger}}{f_{c(2)}^2} -\frac{13}{8} g_2^2 \log \frac{g_2^2 H H^{\dagger}}{2 f_{c(2)}^2}
\right)
\ee
\end{widetext}
where we denote
\bea
\label{defpara}
\mathcal{A}_2 &=& -\frac{1}{f_{c(2)}^2}
\int \frac{d^4 p}{ (2 \pi)^4}   \frac{\delta_{f_{c(2)}} C_{hid}^{(2)} }{p^2}\, \nonumber \\ 
\mathcal{B}_2 &=& -\frac{1}{f_{c(2)}^2} \int \frac{d^4 p}{ (2 \pi)^4}  \frac{\delta_{f_{c(2)}} C_{hid}^{(2)}}{p^2}  \log\frac{f_{c(2)}^2}{p^2} \, .
\eea
Note that $C_{hid}$ are at least one loop, hence $\mathcal{A}_2$
and $\mathcal{B}_2$ are effectively at least at two loop, and they are dimensionless.
An analogous computation for $SU(3)$ leads to
\be
\label{effV3}
\delta_{f_c} V_{eff}^{(3)}=\frac{24 g_3^4 \mathcal{A}_3 }{16 \pi^2}  
f_{c(3)}^2
H H^{\dagger}
\left( -6 \lambda_t^2
\right)\, ,
\ee
with $\mathcal{A}_3$ defined as in (\ref{defpara}) but with $f_{c(3)}^2$ and
$C_{hid}^{(3)}$.

The total effective potential is then the sum of (\ref{effV2}) and (\ref{effV3}). 
If there is not a huge hierarchy between $\mathcal{A}_{2}$ and $\mathcal{B}_2$, one can neglect the 
 terms of order $g_2^4 \lambda$ and $g_2^6$ in (\ref{effV2}) compared to $\lambda_t^2$
 (one can check that including these terms does not affect the discussion). 
 On the other hand we must retain the terms of this order multiplying the  logarithm
of $H H^{\dagger}$ which can become large at the minimum. 

The complete potential then simplifies to
\begin{widetext}
\be
\label{Veff_fin}
V_{eff}= \lambda (H H^{\dagger})^2+
\frac{9 g_2^4 \mathcal{A}_2 }{16 \pi^2}  f_{c(2)}^2  H H^{\dagger}
\left( 4 \pi^2  - \lambda_t^2(6+16 \frac{  f_{c(3)}^2 \mathcal{A}_3}{  f_{c(2)}^2  \mathcal{A}_2} \frac{g_3^4}{g_2^4})+
(\lambda-\frac{13}{8} g_2^2)\log \frac{H H^{\dagger}}{f_c^2}
\right) \, .
\ee
\end{widetext}
This is the generic effective potential for the Higgs in models of GMESB. It contains
two arbitrary scales, $f_{c(2)}$ and 
$f_{c(3)}$, and two dimensionless quantities, $\mathcal{A}_2$
and $\mathcal{A}_3$, which are entirely specified by the hidden sectors to which the gauge bosons of  
$SU(2)$ and $SU(3)$ couple. Even in strongly coupled cases they are given simply by the two point functions 
of the hidden sector currents that break scale invariance.

The potential of eq.(\ref{Veff_fin}) can be specified in terms of 
one dimensionful and one dimensionless combination of parameters,
and it can also be simplified in terms of physical scales.
Indeed, we can rewrite the potential by parameterizing the Higgs field as 
$H={e^{i\xi.\tau} } \left( \begin{array}{c} 0 \nonumber \\ {\phi}/{\sqrt{2}} \end{array} \right) $.
Imposing the minimisation condition at $\phi= \langle \phi \rangle$, 
and trading the unknown parameters $ \mathcal{A}_2  f_{c(2)}^2$ for the second derivatives of the potential around $\phi=\langle \phi \rangle$, we arrive at the generic GMESB Higgs potential;
\be
\label{Vsuper0}
V=
\frac{\lambda}{4} \phi^4+
\frac{1}{4} \phi^2  \left(-m_h^2 +\left(m_h^2-2 \langle \phi  \rangle^2 \lambda \right) \log \left[\frac{\phi^2}{\langle \phi \rangle^2}\right]\right)\, .
\ee
The original parameters are related to $\lambda$ and $m_h$ as 
\bea
\label{cond_min0}
 \hspace{1cm} Y &=& 13 g_2^2-8\lambda \, \,  ; \, \,\, \, X=  \frac{2 \langle \phi \rangle^2 \lambda}{ m_h^2}-1 
\nonumber \\
\mathcal{A}_2 &=& \frac{64 \pi^2}{9 g_2^4 }  \frac{X}{ Y  }  \frac{m_h^2}{ f_{c(2)}^2} \nonumber \\
\mathcal{A}_3 &=&  \frac{\pi^2 }{ 18 g_3^4 \lambda_t^2} 
\mbox{\Huge (}
1
+16 \frac{X}{Y}(2 \pi^2-3 \lambda_t^2)
\nonumber \\ && \hspace{3cm} - \frac{X}{\pi^2}      
\log \frac{\langle \phi \rangle^2}{2 f_{c(2)}^2} \mbox{\Huge )}
  \frac{m_h^2}{ f_{c(3)}^2}\, .
\nonumber
\eea
Since $\mathcal{A}_2$ is generically two loop suppressed, eq.(\ref{cond_min0}) implies that, in the 
absence of any fine-tuning, the scale of new physics is larger than the electroweak scale
by a factor of two to three loops. Hence we have a generic prediction that 
\begin{equation}
\label{fcexpected}
f_c \sim 10^3 -10 ^6 \mbox{~GeV}. \end{equation}
It is interesting to note that the value of $\lambda$ gives direct information about the hidden sector. 
Taking into account that $\frac{m_h^2}{v^2} \simeq \frac{1}{4}$, 
eq.(\ref{cond_min0})
 implies that if $\lambda < \frac{m_h^2}{2 v^2}\simeq \frac{1}{8}$ then the factor $\mathcal{A}_2$
should be negative in order to have electroweak symmetry breaking.
If $ \frac{m_h^2}{2 v^2} < \lambda < \frac{13}{8} g_2^2 $ then it should be positive,
and if $\lambda > \frac{13}{8} g_2^2$ it should again be negative.
An interesting observation is that the SM value, i.e. $\lambda=\frac{m_h^2}{2 \langle \phi \rangle^2}$, is
realised for $\mathcal{A}_2=0$ and $\mathcal{A}_3=\frac{\pi^2 m_h^2}{18  f_{c(3)}^2 g_3^4 \lambda_t^2}$;
in other words the canonical SM potential results when conformal invariance in the hidden sector is 
communicated dominantly by states charged under $SU(3)$.

In the next subsection we discuss the minimal implementation of GMESB, which has
positive $\mathcal{A}_2$ as a natural outcome. This 
leads to correct electroweak symmetry breaking when the quartic
coupling lies in the window $ \frac{m_h^2}{2 v^2} < \lambda < \frac{13}{8} g_2^2 $.
However it is clear from the above that
the generic prediction of GMESB is that the Higgs potential 
takes the form in eq.(\ref{Vsuper0}) regardless of the details of the hidden sector, and that 
the tree level Higgs quartic coupling at the fixed point, $\lambda$, encapsulates the entire difference between this and the Standard Model.

In section \ref{pheno} we will discuss the phenomenology of the
potential  (\ref{Vsuper0}) in the different regimes.

We conclude this discussion with a remark concerning the effective potential (\ref{Vsuper0}).
The logarithmic/quadratic terms in the effective potential deserve comment as they may be 
surprising to readers familiar with perturbative computations and dilaton effective actions. 
They result from the fact that UV scale invariance means that the leading contributions to the quadratic term must  
come from higher order radiative corrections that mix UV (where $f_c$ resides) and IR (where $\phi$ resides) contributions~\footnote{Note that the situation is different from the one described in ref.\cite{gildener} because in that case the two scales $\phi$ and $f_c$ appear at one-loop as the overall ``dilaton'' $\sqrt{f_c^2+\phi^2}$. Here such one-loop terms are necessarily absent.}.

%

We can in fact verify that the method of computation above gives the correct answer in a much simpler (two loop) 
example than the one studied in this paper, namely a theory with two scalars coupled with a quartic coupling, where one 
scalar has a large mass $f_c$. The mass $f_c$ is playing the role of the order parameter for the spontaneous breaking of scale invariance,
corresponding to the vev of the would be dilaton. The effective potential at two loop order can be computed directly as in \cite{Martin:2001vx}, and 
one finds terms going as $ \lambda \lambda_{12} \phi^2 f_c^2 \log \phi^2$, where $\phi$ is the tree-level massless field,
$\lambda_{12}$ is the quartic coupling between the two scalars, and $\lambda$ is the quartic coupling for $\phi$.
The other dimensionful argument of the $\log$ is typically the renormalization scale. One can now realise scale invariance 
non-linearly by using the prescription of ref.\cite{Shaposhnikov:2008xi}, in which one replaces the renormalisation scale with the dilaton.
This results in a term going like $\log \phi^2/f_c^2$. The same terms are obtained if one uses the approach we adopted above for the derivation of eq.\eqref{Veff_fin}. 
Thus we argue that 
our results would be consistent with those derived by performing the entire three-loop computation of the effective action in the theory under study.

\subsection{Minimal GMESB}

In this subsection we introduce a minimal model of GMESB in which the hidden sector is constrained by certain assumptions. 
We will see how the constraint of UV scale invariance allows one to extract quantitative results for $C_{hid}$, and 
hence determine the parameters ${\cal {A}}_{2,3}$ governing the Higgs potential completely.

The minimal model is defined as follows. First, we will assume that the visible sector contains only the SM.  
Second, we assume that the hidden sector 
contains at the fixed point bosonic and fermionic states with unknown anomalous dimensions, induced by the hidden sector dynamics
(which could be strongly coupled), and we will take the anomalous
dimensions of multiplets for each group to be degenerate, $\gamma_F$ for the fermions and $\gamma_B$ for the bosons, with e.g. 
$\text{dim}(\phi)=1+ \gamma_B$. Note that we will take there to be two independent hidden sectors, one coupled to  $SU(3)$ 
and one to $SU(2)$, each determined by their content of bosonic and
fermionic matter. (One could in principle allow states that are charged under $SU(3)$ and $SU(2)$ simultaneously.)
We will also assume that all the hidden sector states have universal mass \footnote{As in the previous section, a simple generalisation would be to take different mass scales $f_c^{(2)}$ and $f_c^{(3)}$ for the different gauge 
groups. In addition if the hidden sector is strongly coupled, one might expect a tower of resonances separated by an interval $f_c$ rather than a universal mass. This could presumably be modelled using bulk propagators for gauge bosons in RS1. We leave these possibilities for future work.}, $f_c$. It will turn out that for consistency $\gamma_F$ and $\gamma_B$ must be negative. 

Whatever the hidden sector configuration, once the matter content is known, the function $C_{hid}$ can in principle 
be expanded perturbatively in the SM gauge couplings. We will here consider only the one loop contribution to $C$. In order to do this, let us first recall the canonical weakly coupled expression for an $SU(N)$ gauge theory with generic matter content of massive scalars and fermions, together with its asymptotics (derived in  Appendix A). Schematically it can be written as 
\bea
C^{1}&=&\frac{1}{8 \pi^2} \mbox{\LARGE (}
\sum_{\phi} C(r_{\phi}) C^1_B(m_{\phi})\nonumber \\
&& \hspace{0.3cm} +\sum_{\psi} C(r_{\psi}) C^1_F(m_{\psi}) +\sum_{G} C(adj) C^1_G\mbox{\LARGE )}
\eea
where $C^1_B(m_{\phi}),C^1_F (m_{\psi}),C^1_G$ denote the one loop contribution to the gauge boson propagator
from complex bosons, Weyl fermions and gauge bosons\footnote{That we consider massless for simplicity.}
running in the loop respectively and 
(at the risk of confusion) $C(r)$ represents the Casimir index of the representation $r$.
The asymptotic expressions are at large momentum (with respect to the characteristic mass scale of the field):
\begin{eqnarray}
C^1_B(m_{\phi)})&=&\frac{1}{3} \log (\frac{\mu^2}{p^2})+2\frac{m_{\phi}^2}{p^2} (1+\log(\frac{m_{\phi}^2}{p^2})) +O(1/p^{4}) , 
\nonumber \\
C^1_F (m_{\psi}) &=& \frac{2}{3} \log(\frac{\mu^2}{p^2})-4 \frac{m_{\psi}^2}{p^2}+O(1/p^{4})\, ,
\end{eqnarray}
and at low momentum:
\bea
C^1_B(m_{\phi)})&=&-\frac{1}{3} \log (\frac{m_{\phi}^2}{\mu^2}) +O(p^{2}) \nonumber \\
C^1_F (m_{\psi}) &=&-\frac{2}{3} \log(\frac{m_{\psi}^2}{\mu^2})+O(p^{2})\, ,
\eea
where we denote with $m_{\phi}$ and $m_{\psi}$ the mass of the scalar and of the fermion,
respectively.

More generally the anomalous dimensions of the hidden sector matter determine the structure of the $C_{hid}$ function and one can argue (as in e.g. \cite{late7}) that the generic expression is obtained by simply modifying the canonical one,
such that all dimensionful quantities scale with appropriate exponents (equivalently that propagators scale as $1/(p^2-m^2)^{1-\gamma_B}$).
 Under these hypotheses, the large and small momentum expansions for $C_{hid}$ are respectively
\begin{widetext}
\bea
C^1_{hid} & \rightarrow & \frac{\delta^{ab}}{8 \pi^2} 
\left[
\log \frac{\mu^2}{p^2}
 \left(
\sum_{\phi} (1-\gamma_B) \frac{C(r_{\phi})}{3} +\sum_{\psi}(1-\gamma_F) \frac{2 C(r_{\psi})}{3}
\right)
\right. 
+\, 2 \frac{f_c^2}{p^2}  \left(
 \frac{f_c^{-2 \gamma_B}}{p^{-2 \gamma_B}}  \sum_{\phi} C(r_{\phi})-2\frac{f_c^{-2 \gamma_F}}{p^{-2 \gamma_F}}  \sum_{\psi}  C(r_{\psi})
\right)
\nonumber \\
&& \label{Chidlarge}
\hspace{7cm}\left.
+\, 2 \frac{f_c^{2(1- \gamma_B)}}{p^{2(1- \gamma_B)}}(1-\gamma_B) \log \frac{f_c^2}{p^2} \sum_{\phi} C(r_{\phi})
\right] +{\mathcal O}(1/p^4)\, ,
\label{Chidlow}
\\
C^1_{hid}
&\rightarrow & -\frac{\delta^{ab}}{8 \pi^2}
\log \frac{f_c^2}{\mu^2}
 \left(
\sum_{\phi} (1-\gamma_B) \frac{C(r_{\phi})}{3} +\sum_{\psi}(1-\gamma_F) \frac{2 C(r_{\psi})}{3}
\right)  +{\mathcal O}(p^2) \, .
\eea
\end{widetext}

The terms proportional to $\log(\mu)$ can be 
recognised as the modification to the
beta function of the gauge coupling coming from fields with large anomalous dimensions; we denote the contribution to the beta function coefficient by
\be
b'_{HS} \equiv \left(
\sum_{\phi} (1-\gamma_B) \frac{C(r_{\phi})}{3} +\sum_{\psi}(1-\gamma_F) \frac{2 C(r_{\psi})}{3}
\right) \, .
\ee
The assumption that conformal invariance is spontaneously broken only 
by $f_c$ imposes some constraints on the asymptotics of the $C$ functions.
In particular, the beta function for the gauge boson should be zero at large momentum.
The beta function is a combination of the wave function renormalisation for the gauge boson,
encapsulated by the asymptotic logarithm of $C_{tot}=C_{vis}+C_{hid}$, the vertex renormalisation and the
wave function renormalisation of matter.
In order for the beta function of the gauge coupling to vanish in the UV, the asymptotic 
behaviour of $C_{tot}$ should be such that
\footnote{We are working in unitary gauge.}
\footnote{{Note that $2 N_c-b'_{SM}=b_0^{SM}$ where $b_0^{SM}$ is the SM one loop beta function coefficient.}}
\be
\label{conf_ass}
 b'_{HS}
+
{b'}_{SM} \simeq 2 N_c\, .
\ee
This should be valid both for the $SU(2)$ and the $SU(3)$ gauge groups.
We write an approximate equality since the two loop contributions could be relevant in determining the fixed point; 
however even in that case
the above relation holds approximately.

We can use the above asymptotics to compute the integrals 
(\ref{defpara}) determining the effective potential, by splitting the domain of integration
into the regions $p^2 \ll f_c^2$ and $p^2 \gg f_c^2$.
We use a shorthand notation for the effective number of fermions and bosons in the 
hidden sector
\be
n_B=\sum_{\phi} C(r_{\phi}) \qquad n_F=\sum_{\psi}  C(r_{\psi})\, .
\ee 
The anomalous dimensions of the hidden sector fields should be negative, in order for the 
integrals to be convergent. Including for completeness  the integral $\mathcal{B}_2$,
we obtain 
\bea
\label{compact1}
&&
\hspace{-0.3cm} \mathcal{A}_a=\frac{1}{(16 \pi^2)^2}
\left(
2 \left( b'_{HS} \right)_{(a)}+\frac{4 n_B^{(a)}}{(\gamma_B^{(a)})^2}-\frac{8 n_F^{(a)}}{\gamma_F^{(a)}}
\right)\, 
\\
&& 
\hspace{-0.3cm} \mathcal{B}_2 \hspace{-0.1mm}= \hspace{-0.1mm}
\frac{1}{(16 \pi^2)^2}\left(
4 \left(b'_{HS} \right)_{(2)}+\frac{4 n_B^{(2)} (2-\gamma_B^{(2)})}{(\gamma_B^{(2)})^3}-\frac{8 n_F^{(2)}}{(\gamma_F^{(2)})^2}
\right)\nonumber 
\eea
where $a=2,3$.
We note that, unless $\frac{1}{\gamma} \gg 1$, there is not a large hierarchy between $\mathcal{A}_2$
and $\mathcal{B}_2$, so we can consistently neglect the terms multiplying $\mathcal{B}_2$ as 
anticipated. The complete potential is then as in (\ref{Veff_fin}) with $\mathcal{A}_2$ and $\mathcal{A}_3$
given by equation (\ref{compact1}).
As expected the quantities $\mathcal{A}_a$ and $\mathcal{B}_2$
are two loop suppressed. 
Moreover, the relation (\ref{conf_ass}) implies that $\left( b'_{HS} \right)_{(a)}$ are positive, both for
$SU(2)$ and $SU(3)$ gauge groups (they are respectively $\left( b'_{HS} \right)_{(2)} \simeq \frac{19}{6} $ and $\left( b'_{HS} \right)_{(3)} \simeq 7$). Since the anomalous dimensions are negative, the factors $\mathcal{A}_a$ are positive
definite. This is a peculiarity of the minimal model and is not a generic property of GMESB models.
However it implies that in the minimal model we cannot realise scenarios with purely logarithmic electroweak
symmetry breaking and negligible quartic coupling (see the discussion at the end of the previous section).

In minimal GMESB, the expected scale of new physics and the precise value of the quartic coupling
clearly depend sensitively on the field content of the hidden sector and on the 
anomalous dimensions, since they determine the
integrals (\ref{defpara}). 

Let us consider a simple example to demonstrate.
Suppose that 
the $SU(2)$ and the $SU(3)$ hidden sectors contain $4$ and $8$ 
pairs of weyl fermions in the fundamental and antifundamental
representation of the gauge group, respectively, so $n_{F(2)}=4$ and $n_{F(3)}=8$.
Both hidden sectors do not contain bosons for simplicity, i.e. $n_{B(2)}=n_{B(3)}=0$. 
Then we can consider two cases for the anomalous dimensions of the fermions in the two hidden sectors:
the first  has $\gamma_F^{(2)}=-0.034$ and $\gamma_F^{(3)}=-0.26$,
the second $\gamma_F^{(2)}=-0.023$ and $\gamma_F^{(3)}=-0.20$.
Note that the fermionic matter content of each hidden sector is such that eq.(\ref{conf_ass}) is roughly satisfied.
Imposing eqs.(\ref{cond_min0}) and demanding the correct $\langle \phi \rangle$ and $m_h^2$, 
the resulting values for the quartic coupling and the scale $f_c$ are
\bea
\gamma_F^{(2)}=-0.034 \, ,  \, \gamma_F^{(3)}=-0.26   \, &\Rightarrow&  \, \lambda=0.15   \, ,  \,  f_c=2.5 \, \text{TeV}\nonumber \\
\gamma_F^{(2)}=-0.023  \,,  \, \gamma_F^{(3)}=-0.20  \, &\Rightarrow&  \, \lambda=0.5  \, ,  \, f_c=15 \, \text{TeV} \nonumber .
\eea
In the first case,
since $\lambda$ is close to the limiting SM value of $\frac{1}{8}$, there has to be a partial cancellation in the
factor $ X=  \frac{2 \langle \phi \rangle^2 \lambda}{ m_h^2}-1$ in equation (\ref{cond_min0}),
leading to a scale of new physics $f_c$ that is at the lower end of the expected window in eq.(\ref{fcexpected}).
In the second case, the value of $\lambda$ is larger than the SM one, there is no
cancellation in equation (\ref{cond_min0}), and the resulting scale of new physics is 
correspondingly larger.

Clearly since $\mathcal{A}_2$ is positive definite in minimal GMESB,
equation (\ref{cond_min0}) tells us that $\lambda$ will always lie between the SM value and $\frac{13}{8} g_2^2$ in this case,
as demonstrated in the above examples. The lower limit yields a potential for the Higgs essentially identical to the SM one, 
and new physics at a relative accessible scale. The upper limit yields a Higgs potential with different self-couplings from the SM potential, but the scale of new physics is pushed beyond the reach of LHC. It is interesting that either limit offers the prospect of experimental detection. 
In section \ref{pheno} we will describe the phenomenological properties of the Higgs potential in the minimal case as well as the other 
\emph{non}-minimal cases described in section~\ref{GMESB}.

\section{The global picture: a toy model}

We wish to briefly discuss the global picture in more detail. In particular
an important issue that we would like to address is the nature of the fixed points. 
It is useful to have a specific model in mind for this question. We shall here consider a pair of coupled QCD theories, one standing
for the visible sector and one for the hidden. 

First consider a single
$SU(N)$ QCD with $F$ flavours of (Dirac) quarks, a singlet scalar
playing the role of the SM Higgs, and a large degenerate Yukawa coupling,
$y$, for $F'\leq F$ of the pairs. These $F'$ pairs represent the visible sector 
matter fields whereas the $F-F'$ quarks that do not couple to the Higgs are messengers; 
in a more realistic model one would among other things 
probably wish to introduce Yukawa hierarchies. The Lagrangian
can be written 
\begin{equation}
\mathcal{L}=-\frac{1}{4g^{2}}FF+i\bar{\psi}\gamma.D\psi+\frac{1}{2}\left(\partial h\right)^{2}-\left(y\bar{t}th+h..c.\right)-\lambda h^{4},
\end{equation}
where the $F'$ quarks with degenerate Yukawa coupling are labelled
$t$. Let $a_{g}=g^{2}/(16\pi^{2})$, $a_{y}=y^{2}/(16\pi^{2})$ and
$a_{\lambda}=\lambda/(16\pi^{2})$. Then we obtain to first order
in yukawa and quartic coupling, and 2-loop order in gauge coupling
the following beta functions;
\begin{eqnarray}
\beta_{\alpha_{g}} & = & -2\alpha_{g}^{2}N\left[\frac{11}{3}-\frac{2}{3}\frac{F}{N}+\alpha_{y}\frac{F'}{N}\right]\nonumber \\
 &  & \,\,\,\,\,\,\,\,\,\,\,\,\,-2a_{g}^{3}N^{2}\left[\frac{34}{3}-\frac{F}{N}\left(\frac{13}{3}-\frac{1}{N^{2}}\right)\right]\nonumber \\
\beta_{\alpha_{y}} & = & 2\alpha_{y}\left[4F'\alpha_{y}-3\frac{N^{2}-1}{N}\alpha_{g}\right]\nonumber \\
\beta_{\alpha_{\lambda}} & = & 2\left[10\alpha_{\lambda}^{2}+2NF'\alpha_{\lambda}\alpha_{y}-NF'\alpha_{y}^{2}\right].\label{eq:rges}
\end{eqnarray}
The canonical fixed point is the Caswell-Banks-Zaks (CBZ) fixed point \cite{Caswell,BZ}, when the number of flavours is chosen so as to make the
one loop contribution to $\beta_{\alpha_{g}}\approx0$. That is, writing
$\beta_{\alpha_{g}}=-\frac{2}{3}\alpha_{g}^{2}N\left[b_{0}+b_{1}\alpha_{g}+b_{y}\alpha_{y}\right]$
we require $b_{0}\approx0$. In order to remain perturbative while
still having the two-loop contribution compensate for the one-loop
contribution we then require $F=\frac{11N}{2}\left(1-\epsilon\right)$,
with $b_{0}=\epsilon\sim1/N$. The usual CBZ fixed point is an asymptotically
free theory ($b_{0}>0$) with a perturbative IR fixed point which
requires $1\gg\epsilon>0$. Perturbativity is still possible because,
since $b_{1}\sim N$, then $b_{1}\alpha_{g}\sim b_{0}$ implies that
the 't~Hooft coupling $4\pi N\alpha_{g}\sim\epsilon\sim1/N$ can
still be much less than one. 
This type of fixed point is a natural
place to consider putting our conformally extended SM. As long as it is IR stable,
the theory does not stray from the fixed point until it is disturbed
by $f_{c}$ sized deformations. As we shall see the latter are naturally
$f_{c}$ sized mass terms for the $F-F'$ (messenger) quarks that do not couple
to the $\phi$ in Yukawa couplings. 

In addition we would like the SM to remain perturbative
so that the fixed-point values of the gauge couplings are not dramatically
different from their usual SM values.

To leading order in $1/N$, one can find a non-trivial fixed point
at
\begin{equation}
4\pi N\left\{ \alpha_{g*},\,\alpha_{y*},\,\alpha_{\kappa*}\right\} =4\pi\frac{11\epsilon}{50}\left\{ \frac{4}{3},\,\frac{N}{F'},\,\frac{N}{2F'}\right\} .\label{eq:fps}
\end{equation}
Hence we require $\epsilon>0$, but clearly the couplings can be arbitrarily
small at the fixed point provided that $F'\gtrsim N$. Defining $\gamma_{i}=\frac{\alpha_{i}}{\alpha_{i}^{*}}-1$,
and diagonalizing the linearised RGEs in (\ref{eq:rges}) about the
fixed points, the beta functions have the eigenvalues
\begin{equation}
\beta_{\gamma_{i}}=4\frac{11}{25}\epsilon\left\{ \frac{11}{9}\epsilon,\,\frac{N}{2},\,1\right\} ,
\end{equation}
and, since $\epsilon>0$, all eigenvalues are positive indicating
an IR stable fixed point as required. The small eigenvalue corresponds mainly
to the gauge coupling for which the fixed point is only weakly attractive
(so called quasi-fixed behaviour).

What about the hidden sector theory? We need to ensure that this theory
flows away from its fixed point, so that it is able to generate
the spontaneous breaking of scale invariance $f_{c}$ through some
dynamical mechanism, possibly confinement with a mass-gap. This requires
a fixed point that is unstable in the IR. It is possible to show (using
an analysis similar to the one in \cite{Martin:2000cr}) that one cannot
have a perturbative UV fixed point in which all directions are UV
stable. Essentially the issue is that, as can be seen from (\ref{eq:rges}),
when $b_{0}\approx0$ (for perturbativity) then $b_{1}<0$, whereas
a UV stable fixed point requires $b_{1}>0$. In fact this is true
in general for any additional states that one adds. However (unlike
\cite{Martin:2000cr} which was concerned with solving Landau pole
problems) here we only require that \emph{some }directions are
unstable in the IR in order to start a flow. 

Consider therefore $SU(N)$ QCD with $F''$ flavours of (Dirac) quarks,
$f$, with $SU(F'')_{L}\times SU(F'')_{R}$ flavour symmetry, and a Yukawa
coupling $Y$ to a flavour bi-fundamental scalar, $\Phi$. The Lagrangian
can be written 
\bea
\mathcal{L} &=& -\frac{1}{4g^{2}}FF+i\bar{f}\gamma.Df+\frac{1}{2}\left(\partial\Phi\right)^{2}-\left(Y\bar{f}f\Phi+h..c.\right)\nonumber \\
&& \qquad \qquad \qquad -\lambda\left[\mbox{Tr}(\Phi^{\dagger}\Phi)\right]^{2}-\kappa\mbox{Tr}\left[(\Phi^{\dagger}\Phi)^{2}\right].
\eea
 Note that similar systems were analysed in \cite{Antipin:2013pya}. 
 To 1-loop order in the yukawa and quartic coupling, and 2-loop order
in the gauge coupling, the beta functions are
\begin{eqnarray}
\beta_{\alpha_{g}} & = & -2\alpha_{g}^{2}N\left[\frac{11}{3}-\frac{2}{3}\frac{F''}{N}+\alpha_{y}\frac{F^{\prime\prime 2}}{N}\right]\nonumber \\
 &  & \,\,\,\,\,\,\,\,\,\,\,\,\,-2a_{g}^{3}N^{2}\left[\frac{34}{3}-\frac{F''}{N}\left(\frac{13}{3}-\frac{1}{N^{2}}\right)\right]\nonumber \\
\beta_{\alpha_{Y}} & = & 2\alpha_{g}\left[(F''+N)\alpha_{Y}-3\frac{N^{2}-1}{N}\alpha_{g}\right]\nonumber \\
\beta_{\alpha_{\lambda}} & = & 2\left[4N\alpha_{\lambda}^{2}+2N\alpha_{\lambda}\alpha_{Y}-F''\alpha_{Y}^{2}\right].\label{eq:rges-1}
\end{eqnarray}

To this order, $\kappa$ runs independently and has a fixed point
at $\kappa=0$.
Again to retain perturbativity we choose $F''=\frac{11}{2}N\left(1-\epsilon\right)$,
with $b_{0}=\epsilon\sim1/N$. To leading order in $1/N$ and $\epsilon$,
there is a fixed point at
\begin{equation}
4\pi N\left\{ \alpha_{g*},\,\alpha_{y*},\,\alpha_{\kappa*}\right\} =-4\pi\frac{22}{19}\epsilon\left\{ 1,\,\frac{13}{6},\,\frac{\sqrt{23}-1}{4}\right\} .
\end{equation}
Therefore now we require $\epsilon<0$ for this to be physical. Since
$b_{0}<0$ as well, one suspects that the fixed point is a saddle
point, and indeed it is. Diagonalizing the RGEs about the fixed points,
the beta functions have the eigenvalues
\begin{equation}
\beta_{\gamma_{i}}=13\frac{22}{19}|\epsilon|\left\{ -\frac{11}{9}|\epsilon|,\,4\frac{\sqrt{23}}{13},\,1\right\} .
\end{equation}
The fixed point is weakly repulsive (as required) in the direction that is mainly
comprised of the gauge coupling.

Now we can consider a global configuration in which two such theories
are coupled. For our SM theory, we focus on $SU(3)_{c}$. (Obviously
in a full model there would have to be some degree of unification
in order to include the full SM gauge groups including hypercharge.)
There are 6 $SU(3)_{c}$ flavours in the SM, and to be near a CBZ
fixed point we require $F\lesssim33/2$. Indeed from eq.(\ref{eq:fps})
it is natural to choose $\frac{88\pi}{225}\epsilon\sim\alpha_{s}$
or $\epsilon\approx0.1$, so that the fixed point value of the $SU(3)_{c}$
coupling is not far from its weak-scale value. We can choose $F=15$
(and hence $\epsilon=1/11$), and therefore need to add $9$ pairs of messenger quarks
with $F-F'=9$. By (\ref{eq:fps}) the couplings at the fixed points are given by
\begin{equation}
4\pi\left\{ \alpha_{g*},\,\alpha_{y*},\,\alpha_{\kappa*}\right\} =\frac{2\pi}{25}\left\{ \frac{4}{3},\,\frac{1}{5},\,\frac{1}{10}\right\} ,
\end{equation}
so the theory is just as perturbative as the SM there. Note that the yukawa
couplings are $y\approx0.8$ and the quartic coupling is $\lambda\approx0.3$.
These are reasonable values, for what is after all just a toy model. 

Meanwhile the 9 pairs of quarks have an $SU(9)_{L}\times SU(9)_{R}$
flavour symmetry. We gauge the diagonal anomaly-free part, to be our
hidden sector gauge group. Thus the hidden sector already sees 3 pairs
of messenger quarks (since the bi-fundamentals are charged under $SU(3)_{c}$),
and as we saw above we need to make this up to $F''=\frac{11}{2}N(1-\epsilon)$
where $N=9$, with $\epsilon<0$. Here the solution can be virtually
as weakly coupled as desired, since when $F''=50$ one has $\epsilon=1/99$.
Since GMESB works best with large anomalous dimensions in the hidden
sector, relatively strong coupling is perhaps more desirable however. This may also be required for the
dynamics that leads to the spontaneously breaking of scale invariance. Although it is 
not obligatory, the minimal and perturbative choice is that the field $\Phi$ is the dilaton that gets a vev $f_c$. 
We should stress however that the hidden sector gauge group may become strongly coupled in the 
IR in which case the mediation may be better described by a low energy theory of bound states. This would of 
course greatly complicate the explicit computation of the parameters ${\cal A}_{2,3}$. The
particle content is summarised in Table \ref{tab:Model-table}. 

\begin{table}
\noindent \begin{centering}
\begin{tabular}{|c|c|c|c|}
\hline 
 & $SU(3)_{c}$ & $SU(9)_{L}$ & $SU(9)_{R}$\tabularnewline
\hline 
\hline 
$6\times\psi_{SM}$ & $\square$ & $1$ & $1$\tabularnewline
\hline 
$6\times\bar{\psi}_{SM}$ & $\overline{\square}$ & $1$ & $1$\tabularnewline
\hline 
$\phi$ & $1$ & $1$ & $1$\tabularnewline
\hline 
\hline 
$\psi_{mess}$ & $\square$ & $\overline{\square}$ & $1$\tabularnewline
\hline 
$\bar{\psi}_{mess}$ & $\overline{\square}$ & $ $$1$ & $\square$\tabularnewline
\hline 
\hline 
$(F''-3)\times\psi_{hid}$ & $1$ & $\overline{\square}$ & $1$\tabularnewline
\hline 
$(F''-3)\times\bar{\psi}_{hid}$ & $1$ & $1$ & $\square$\tabularnewline
\hline 
$\Phi$ & $1$ & $\square$ & $\overline{\square}$\tabularnewline
\hline 
\end{tabular}\caption{\label{tab:Model-table}A toy model for GMESB. The $SU(3)_{c}$ theory
sits at an IR fixed point while the $SU(9)_{hid}=\left[SU(9)_{L}\times SU(9)_{R}\right]_{diag}$
theory flows from a repulsive UV fixed point. The $F''$ pairs of
$SU(9)_{hid}$ fermions $f\equiv\{\psi_{mess},\psi_{hid}\}$ and $\bar{f}\equiv\{\bar{\psi}_{mess},\bar{\psi}_{hid}\}$
include the bifundamental messenger fields. }

\par\end{centering}

\end{table}

\section{Summary and phenomenological discussion}\label{pheno}

Let us recap what we have found. We assumed that the Standard Model emanates from a UV scale invariant theory that also contains a messenger sector of fields charged under the SM gauge groups but without couplings to the Higgs. If the messenger fields gain masses of order $f_c$ from spontaneous scale breaking in the rest of the hidden sector, then 
mass terms and
 novel logarithmic couplings are induced radiatively in the Higgs potential. 
 For this to be phenomenologically viable, the scale of new physics should be order $1 - 10^3$~TeV, at which scale one would expect a large number of new states charged under the Standard Model gauge group to appear. Nevertheless there are interesting footprints in the Standard Model, in particular the Higgs self-couplings.

Indeed, the effective potential of the SM Higgs in the IR theory was found to be 
\be
\label{Vsuper00}
V=
\frac{\lambda}{4} \phi^4+
\frac{1}{4} \phi^2  \left(-m_h^2 +\left(m_h^2-2 \langle \phi  \rangle^2 \lambda \right) \log \left[\frac{\phi^2}{\langle \phi \rangle^2}\right]\right)\, .
\ee
The phenomenology depends entirely on how close the quartic coupling $\lambda$ is to its SM value $\lambda=m_h^2/2\langle \phi\rangle ^2 $. This is in turn related 
via eq.(\ref{cond_min0}) to the shift in the vacuum polarisation of the gauge bosons
communicated by the messenger fields,
\begin{equation}
\mathcal{A}_a=\frac{1}{f_{c(a)}^2}
\int \frac{d^4 p}{ (2 \pi)^4}\frac{1}{p^2}    \left( C_{hid}^{(a)}(p^2, 0) -  C_{hid}^{(a)}(p^2, f_c^2)\right)\, ,
\end{equation}
where $a$ labels the gauge group. We stress that these quantities and hence $\lambda $ are calculable in many specific models. We computed it for the minimal case of a perturbative hidden sector, but there are many alternative possibilities. These may include unification at the UV fixed point, and as mentioned may involve strong coupling (in which case one would naturally use RS1 with $f_c$ corresponding to the position of the IR brane, and the $C$'s being given by the propagators of bulk gauge fields). 

\begin{figure}[ht]
\begin{center}
\includegraphics[width=6.5cm]{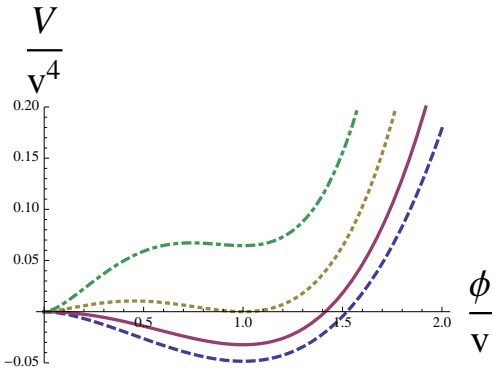}
\caption{
\label{Vsimplplot}\footnotesize
Higgs potential with $m_h^2=125^2$, $\langle \phi \rangle\equiv v=246$ and $\lambda=(\frac{1}{4},\frac{1}{2},1,2) \frac{m_h^2}{\langle \phi \rangle^2}$.
}
\end{center}
\end{figure}

We can identify three  generic phenomenological patterns. These are represented in figure~\ref{Vsimplplot} which shows the potential for the four prototypical values,  $\lambda=(\frac{1}{4},\frac{1}{2},1,2) \frac{m_h^2}{\langle \phi \rangle^2}$. We shall now briefly describe them. \\

\paragraph{\bf Degenerate or metastable electroweak symmetry breaking:}
These configurations are attained for 
$\lambda \geq \frac{m_h^2}{\langle \phi \rangle^2} \simeq \frac{1}{4} $.
The green and yellow lines correspond to $\lambda= 2 \frac{m_h^2}{\langle \phi \rangle^2}\simeq \frac{1}{2}$ and 
$\lambda=\frac{m_h^2}{\langle \phi \rangle^2} \simeq \frac{1}{4}$ respectively. Both these values have
$ \frac{m_h^2}{2 \langle \phi \rangle^2} < \lambda < \frac{13}{8} g_2^2 \simeq 0.7 $ so that ${\cal A}_2>0$ and they can therefore be achieved in the minimal model presented in section 2.1. The metastable case shown is relatively short lived, but by choosing values of $\lambda $ closer to the critical value one can give it a lifetime longer than the age of the Universe. It is interesting to derive the resulting limits on $\lambda$ coming from the decay rate from the metastable vacuum. An estimate of the O$_4$ symmetric bounce action is  \cite{Coleman:1977py}
\begin{equation}
S_{O_4}\sim \frac{2\pi^2 \langle \phi \rangle^4}{V(\langle \phi \rangle)-V(0)}\, . 
\end{equation} 
A value of the action $S\gtrsim 400 $ is generally sufficient to prevent vacuum decay within the lifetime of the Universe. Defining 
$\lambda=\frac{m_h^2}{\langle \phi \rangle^2}+\kappa$, the metastable minimum has vacuum energy raised by $V(\langle \phi \rangle)-V(0)= \frac{\kappa}{4} \langle \phi \rangle^4$. Hence the vacuum is metastable but has insignificant decay if $0< \kappa \lesssim \pi^2/50$, or in other words if $ 0.25 < \lambda \lesssim 0.45$. 
Note that it requires almost no fine-tuning of 
$\lambda$ to protect the metastable vacuum. (On the contrary rapid vacuum decay is actually rather hard to achieve.) It would be interesting to investigate the cosmological consequences of such a potential. We leave this as a subject for future study. 

Finally the case with $\lambda=\frac{m_h^2}{\langle \phi \rangle^2}$ leads to two degenerate vacua, both at zero vacuum energy. The expansion of eq.(\ref{Vsuper00}) around the minimum as $\phi = \langle \phi \rangle +h$, (given for the general case in eq.(\ref{Vhiggs})) is 
\be
V=\frac{m_h^2}{2} h^2+\frac{5 m_h^2}{6 \langle \phi \rangle} h^3 + \frac{7 m_h^2}{24 \langle \phi \rangle^2}h^4 
+ {\mathcal O}(h^5)\, .
\ee

\paragraph{\bf Standard Model with Higgs portal:} 
The red line in  figure~\ref{Vsimplplot} has the Standard Model value of the quartic coupling, namely $\lambda= \frac{m_h^2}{2\langle \phi \rangle^2}$. Clearly for this value, the logarithm in the potential (\ref{Vsuper00}) cancels, the quadratic term is negative, and 
the conventional Standard Model picture follows. The GMESB mechanism is then simply providing a UV completion for the usual Higgs portal model, with
the portal coupling to the hidden sector spurion $\langle \chi \rangle=f_c$ being generated radiatively by gauge mediation.
This case can be realised with ${\cal A}_3>0$ and ${\cal A}_2=0$. In other words the Standard Model is a general prediction of gauge mediation of spontaneous scale breaking in which the mediation is dominated by coloured degrees of freedom only. Since ${\cal A}_3$ is positive it may be achieved with a minimal configuration. For comparison, the 
potential expanded around the minimum is
\be
V=-\frac{m_h^2 \langle \phi \rangle^2}{8}+\frac{m_h^2}{2} h^2+\frac{m_h^2 }{2 \langle \phi \rangle}h^3+\frac{m_h^2 }{8 \langle \phi \rangle^2}h^4\, .
\ee

\paragraph{\bf Logarithmic Higgs potential:} 
Finally in figure~\ref{Vsimplplot}, the blue line shows an example where the quartic term is becoming negligible 
and the minimisation is dominated by the logarithm. In the limit that $\lambda$ is negligible in (\ref{Vsuper00}) with the 
potential taking the approximate form, 
\begin{equation}
V=\frac{1}{4}m_{h}^{2}\phi^{2}\left(\log\frac{\phi^{2}}{\left\langle \phi\right\rangle ^{2}}-1\right) \, ,
\end{equation}
the Higgs self couplings are given by 
\be
\label{VhiggsUS}
V=
-\frac{ m_h^2 \langle \phi \rangle^2}{4} +\frac{m_h^2}{2} h^2+\frac{m_h^2}{6 \langle \phi \rangle} h^3-\frac{m_h^2}{24 \langle \phi \rangle^2} h^4+ {\mathcal O}(h^5)\, .
\ee
~~\\

In summary we see that GMESB leads to a more general Higgs potential, that can have unusual deviations from the conventional 
Standard Model self-couplings. In all other respects however, in particular in its couplings to the matter fields and gauge bosons, the Higgs behaves precisely as it would do in the conventional SM. Note that this is markedly different from the usual 
phenomenology of a Higgs-like dilaton (essentially because the Higgs is not actually the real dilaton): in the conventional notation of e.g. ref.\cite{Serra:2013kga}, only the $c_3$ operator is affected. 

It is also interesting to note that, in common with GMSB, the flavour problem would be greatly ameliorated. As we have mentioned, depending on the 
manner in which scale invariance is restored in the UV, there could conceivably be light states of order a few TeV in non-minimal realisations. These states would be expected to receive order TeV masses much like the sfermions in GMSB, but (as in GMSB) these masses would be degenerate because they depend only on the gauge charges of the fields. Thus any mixing angles could only arise from the CKM matrix, and one would expect the GIM mechanism to still be in operation. 

Finally we can estimate the mass of the dilaton in GMESB (identified with the field that spontaneously breaks scale invariance) which is rather heavy. Its mass follows from the usual PCDC relation, $f_c^2 m_\sigma^2
= -\langle T^\mu _\mu \rangle = -16 \epsilon_{vac}$, where $\epsilon_{vac}$ is the contribution to the vacuum energy density
from the hidden sector
 (see for example the discussion in ref.\cite{Coriano:2012nm}). Thus a perturbative hidden sector would have a dilaton with mass 
$m_\sigma \sim f_c / 4\pi \sim 1-10$~TeV, while a non-perturbative one would have  a dilaton with mass 
$m_\sigma \sim f_c  \sim 10-10^3$~TeV.

\vspace{-0.2cm}

\subsection*{Note added in Proof}

\vspace{-0.2cm}

 During the final stages of publication ref.\cite{Elias-Miro:2014pca}
appeared on the ArXiV. The authors of that paper were concerned that the presence of the logarithmic mass-squared terms 
would not be consistent with the decoupling theorem. We argue that these concerns do not apply to the present case, which is
in accord with it. Indeed these radiatively generated terms have the same status as the logarithmic terms in the Coleman Weinberg potential, and should be thought of as radiative corrections to the quartic coupling between the higgs and hidden sector dilaton $f_c$.  
Although at first sight the example discussed in ref.\cite{Elias-Miro:2014pca} resembles the one that we have presented, the crucial difference is that in our scenario $f_c$ cannot generally
be decoupled without also decoupling the higgs. 
The spurion $f_c$ represents the deviation from a scale invariant fixed point so there can be no parametrically large hierarchy of mass scales that would allow it to be decoupled while leaving the Higgs light.
For example, we see this explicitly in the minimal GMESB case where the ratio of the higgs mass to $f_c$ is fixed by the particle content, anomalous dimensions and couplings at the UV fixed point. 
The only hierarchies of mass scales in the present theory are generated by loop factors (much as in Coleman-Weinberg) which is why $f_c$ is constrained 
to be no more than two or three loops larger than the electroweak scale. 
As a final remark we note that {\emph{ in principle}} it is possible to perform a decoupling limit by sending $f_c$ to infinity while keeping the higgs mass constant by adjusting the couplings (which go to zero). One observes that in this limit the logarithmic terms presented in our potential dutifully disappear, but then the situation is of less phenomenological interest and the robustness of the computations are questionable. \\

\paragraph*{Acknowledgements} We would like to thank Riccardo Argurio, James Barnard, Matt Buican, Jos\'e-Ram\'on Espinosa, Zohar Komargodski, Valya Khoze and Diego Redigolo for interesting dicussions. 
S.A.A. would like to thank the CERN theory division,
the Galileo Galilei Institute for Theoretical Physics and the INFN for hospitality and partial support during the final stages of this work.
A.M. acknowledges funding by a Durham International Junior Research Fellowship. We acknowledge funding from the STFC.


\newpage 
\begin{appendices}

\section{Why scale invariance does not usually protect the Higgs mass}

\noindent It is instructive to ask when a restoration of scale invariance in the UV can be sufficient 
to protect the Higgs mass. Here we shall discuss this question by examining a na\"ive perturbative example, in which  
one attempts to restore scale invariance simply through the addition of extra states or resonances.
At the same time this gives the opportunity to present some expressions required for the main text. 

Consider the SM, but turn off all the
Yukawa couplings and $SU(3)_{c}\times U(1)_{Y}$ gauge couplings and
also neglect the Higgs self-coupling so that the $SU(2)_{W}$ coupling
dominates. Now make the theory run perturbatively to a UV fixed
point at a high scale. Since $SU(2)_{W}$ has a non-zero beta function
in the SM this requires extra degrees of freedom, and we can take
them to be simply heavy $SU(2)$ \emph{fermion doublets} $\eta_{L_{i}},\eta_{R_{i}}$,
and \emph{scalar doublets} $\tilde{\eta}_{L_{a}},\tilde{\eta}_{R_{a}}$. 
This general case includes the supersymmetric system which is an interesting
point of reference. As we shall show in this Appendix and in Appendix B, generally such a theory still suffers from the 
hierarchy problem because UV sensitivity is not removed and therefore it is not possible to take the continuum limit, i.e. send the UV cut-off $\Lambda_\infty$ to infinity. We will see that  
UV sensitivity is removed entirely in a perturbative theory when the 
poles of heavy states form a multipole (in momentum-space) of sufficiently high order. Nonperturbatively, and in the Wilsonian language, one would say that the theory has to lie on a 
``renormalisable trajectory'' (i.e. a self-similar RG trajectory that emanates from a repulsive UV fixed point - see for example the review of
 \cite{Rosten:2010vm}).

\noindent Using dimensional regularisation, we can compute the radiative
contributions to the Higgs mass in two stages. First the contribution
to the gauge propagator can be captured in the current-current correlators
for $SU(2)$; 
\begin{equation}
\text{\textlangle}j_{\mu}(p)j_{\nu}(-p)\text{\textrangle}=-(\eta_{\mu\nu}p^{2}-p_{\mu}p_{\nu}){C}(\frac{p^{2}}{\mu^{2}},\epsilon),
\end{equation}
with the effective Lagragian being given by 
\begin{equation}
\delta\mathcal{L}=-\frac{1}{4}g^{2}C F_{\mu\nu}F^{\mu\nu}.
\end{equation}

The subsequent contribution to the potential (in Euclidean coordinates)
is 
\begin{equation}
\delta V=3g^{2}\int\frac{p^{2}d^{4}p}{(2\pi)^{4}p^{2}+m_{W}^{2}}\left(C(\frac{p^{2}}{\mu^{2}},\epsilon)-\Delta_{Z}\right)\label{eq:mh2}
\end{equation}
with $\Delta_{Z}$ being the counter-terms. It is useful to define
parameters scaled by $f_{c}$ with hats, so that
\begin{equation}
\delta V=3g^{2}f_{c}^{4}\int\frac{\hat{p}^{2}d^{4}\hat{p}}{(2\pi)^{4}\hat{p}^{2}+\hat{m}_{W}^{2}}\left(C(\hat{p}^{2},\hat{\mu}^{2},\epsilon)-\Delta_{Z}\right)\label{eq:mh2-1}
\end{equation}
The dimensionless parameter $C$ can be evaluated in general
terms: defining masses scalar and fermion
masses as $M_{a}$ and $M_{i}$ respectively, we find contributions
of the form 
\begin{widetext}
\begin{equation}
i{C}(\hat{p}^{2},\hat{\mu}^{2},\epsilon)=\frac{2g^{2}C_{R}}{3}\left(I(p,M_{a})+2I(p,M_{i})\right)+\frac{8g^{2}C_{R}}{3p{}^{2}}\left[K(p,M_{a})-K(p,M_{i})\right]\label{eq:c1}
\end{equation}
\end{widetext}

where 
\begin{equation}
K(p,m)=m^{2}I(p,m)+J(m)+\frac{im^{2}}{16\pi^{2}}
\end{equation}
and where 
\begin{eqnarray}
I(p,m) & = & i\int\frac{d^{4}q}{(2\pi)^{4}}\frac{1}{(q^{2}+m^{2})((q+p)^{2}+m^{2})}\nonumber \\
 & = & i\frac{1}{16\pi^{2}}\left(\frac{2}{\epsilon}+\Gamma'(1)+\log\left[\frac{4\pi\mu^{2}}{m^{2}}\right]\right. 
 \nonumber \\ && 
 \left. \qquad -\int_{0}^{1}dx\,\log\left[1+\frac{p^{2}}{m^{2}}x(1-x)\right]\right)\\
J(m) & = & i\int\frac{d^{4}q}{(2\pi)^{4}}\frac{1}{(q^{2}+m^{2})} \\
 & = & -i\frac{m^{2}}{16\pi^{2}}\left(\frac{2}{\epsilon}+\Gamma'(1)+1+\log\left[\frac{4\pi\mu^{2}}{m^{2}}\right]\right)\nonumber ,
\end{eqnarray}
so that 
\[
K(p,m)=-i\frac{m^{2}}{16\pi^{2}}\int_{0}^{1}dx\,\log\left[1+\frac{p^{2}}{m^{2}}x(1-x)\right].
\]
Scaling violation is parameterised by $\mu$, which is directly
attributed to RG-flow, whereas infinities appear as poles in $\epsilon$.
In a fully scale invariant theory (i.e. in the deep UV) the total
dependence on $\mu$ should vanish which means that $\epsilon $ poles also cancel. In a cut-off regulated theory (c.f. Appendix B) poles in $\epsilon $ are replaced by 
dependence on the UV cut-off $\Lambda_\infty$. Full scale invariance means that there is no remaining dependence 
on $\Lambda_\infty$, so that one can take the continuum limit, $\Lambda_\infty\rightarrow \infty$, as desired. 
Including the rest of the Standard Model and Higgses, and assuming
only $SU(2)$ fundamentals, we find that the first piece in eq.(\ref{eq:c1})
in conjunction with the obvious gauge loops and one-loop vertex correction, gives the beta function
\begin{eqnarray}
\beta & = & -\frac{g^{3}}{16\pi^{2}}\mbox{\LARGE ( } \frac{11}{3}C_{G}-\frac{4}{3}(N_{\mbox{\tiny{SM-ferm}}}+N_{\eta})C_{\mbox{\tiny{fund}}}\nonumber \\
&& \qquad\qquad \qquad\qquad-\frac{2}{3}(N_{h}+N_{\tilde{\eta}})C_{\mbox{\tiny{fund}}}\mbox{\LARGE )} \nonumber \\
 & = & -\frac{g^{3}}{16\pi^{2}}\left(\frac{11}{3}N_{c}-\frac{2}{3}(N_{\mbox{\tiny{SM-ferm}}}+N_{\eta})\right. \nonumber \\
 && \qquad\qquad\qquad \qquad\qquad \left. -\frac{1}{3}(N_{h}+N_{\tilde{\eta}})\right),
\end{eqnarray}
where the $N$'s count the number of complete Dirac flavours, with
each pair of complex scalars (in a hang-on from supersymmetry) counting
one to $N_{\tilde{\eta}}$. In the SM there are $12$ fermion doublets
($N_{SM-ferm}=6$) and a single Higgs ($N_{h}=1/2$) giving $\beta_{UV}=-\frac{g^{3}}{16\pi^{2}}\left(\frac{19}{6}-\frac{1}{3}(2N_{\eta}+N_{\tilde{\eta}})\right)$.
In order to get conformal behaviour in the UV we could then simply
choose the heavy states such that $(2N_{\eta}+N_{\tilde{\eta}})=3b_{SU(2)}^{(SM)}=\frac{19}{2}$
(noting that scalars contribute $1/2$). 

A useful renormalisation scheme for a theory that flows to a fixed point in the UV is the so-called ``sliding-scale scheme'' \cite{scrucca}
where ${C}_{1}(a,a,\epsilon)-\Delta_{Z}=0$, because one does
not need to perform matching and introduce thresholds by hand. Indeed this renormalisation scheme 
is the one that most closely mimics Wilsonian renormalisation, as it effectively subtracts contributions 
from modes above the RG scale, $\mu$. 
Defining $\zeta(x)=\sqrt{-4/x-1}$, from the above discussion we find that the nett
contribution to the vacuum polarisation of a single boson or Weyl-fermion is 
\begin{eqnarray}
&& \hspace{0.2cm} {C}_{m}(\hat{p}^{2},\hat{\mu}^{2}) =  \frac{g^{2}}{16\pi^{2}}\left[H{}_{\pm}(\hat{p}^{2})-H_{\pm}(\hat{\mu}^{2})\right]\nonumber \\
&& H_{\pm}(z) = -\frac{4}{3}\left(\left(\kappa_{\pm}\pm\frac{2}{z}\right)\zeta\left(z\right)\tan^{-1}\left[\frac{1}{\zeta\left(z\right)}\right]\mp2z^{-1}\right)
\nonumber \end{eqnarray}
where $\mu$ is the RG scale and $\hat{p}=p/m$, $\hat{\mu}=\mu/m$,   
$\pm$ refers to the spin statistics, and where $\kappa_{\pm}=1$
or $\frac{1}{2}$ respectively. The function $H_{\pm}$ has the limiting
behaviour
\begin{eqnarray}
\label{eq:limit-approx}
&&\hspace{-0.3cm}\lim_{z\rightarrow0} H_{\pm}(z)  =  \mbox{const.} \\
&& \hspace{-0.3cm}H_{\pm}(z)  {\stackrel{z\rightarrow\infty}{\longrightarrow} }  
\left\{
\begin{array}{cc}
-\frac{1}{3}\log z-\frac{2}{z}\left(1-\log z\right)+\mathcal{O}(z^{-2}) & \hspace{-0.2cm}\mbox{:boson}  \\
-\frac{2}{3}\log z-\frac{4}{z}+\mathcal{O}(z^{-2}) & \hspace{-0.3cm} \mbox{:fermion.}
\nonumber \end{array}
\right.
\end{eqnarray}
There are then three interesting limits; $\mu,p\gg m$ ($\hat{\mu},\hat{p}\gg1$)
which is when the mass is negligible, $\mu,p\ll m$ ($\hat{\mu},\hat{p}\ll1$)
when the state is heavier than both the momentum and the RG scale,
and $\mu\gg m\gg p$ ($\hat{\mu}\gg1\gg\hat{p}$) which is when the
RG scale is larger than the mass, but the momentum is small. Defining
$\Delta b=-\frac{1}{3},-\frac{2}{3}$ to be the contribution of the
state to the $\beta$-function coefficient, to leading order these
limits give:
\begin{equation}
\label{eq:somelabel}
\frac{16\pi^{2}}{g^{2}}{C}_{m}(\hat{p}^{2},\hat{\mu}^{2})\rightarrow
\left\{
\begin{array}{ll}
\Delta b\,\log\frac{p^{2}}{\mu^{2}} & \mu,p\gg m\\
0 & \mu,p\ll m\\
\Delta b\,\log\frac{m^{2}}{\mu^{2}} & \mu\gg m\gg p.
\end{array}
\right.
\end{equation}
In the present context the contribution to the total polarisation from the SM fields is cancelled in the UV by 
states of mass $f_c$ which for convenience we assume are degenerate (and we also
neglect the SM masses). Therefore we must choose $N_\eta$ and $N_{\tilde{\eta}}$ accordingly (as above) such that the 
total vacuum polarisation including the SM states is 
\begin{widetext} 
{\begin{eqnarray}
\frac{16\pi^{2}}{g^{2}}{C}_{tot}(\hat{p}^{2},\hat{\mu}^{2}) & = & \,b_{SU(2)}^{\prime (SM)} \log\left[\frac{\hat{p}^{2}}{\hat{\mu}^{2}}\right]
-b_{SU(2)}^{(SM)}
\int_{0}^{1}dx \log\left[\frac{1+\hat{p}^{2}x(1-x)}{1+\hat{\mu}^{2}x(1-x)}\right]\nonumber \\
 &  & -\frac{4(N_{\tilde{\eta}}-N_{\eta})}{3}\int_{0}^{1}dx\,\frac{1}{\hat{p}^{2}}\log\left[1+\hat{p}^{2}x(1-x)\right]-\frac{1}{\hat{\mu}^{2}}\log\left[1+\hat{\mu}^{2}x(1-x)\right]
 \label{eq:master}
\end{eqnarray}
}\end{widetext} 
where $b_{SU(2)}^{ (SM)}$ is the total $\beta$-function coefficient 
of the SM, and $b_{SU(2)}^{\prime  (SM)}$ is the $\beta$-function coefficient of the SM coming from vacuum polarisation diagrams 
(there is a piece that can be attributed to the one-loop vertex 
diagram as well). Note that ${C}_{1-tot}(a,a)=0$
which is the sliding-scale condition satisfied
by construction. From (\ref{eq:somelabel}),  the various limits of the vacuum polarisation are 
\begin{widetext}
\begin{equation}
\frac{16\pi^{2}}{g^{2}}{C}_{tot}(\hat{p}^{2},\hat{\mu}^{2})\rightarrow
\left\{
\begin{array}{ll}
\left( b_{SU(2)}^{ \prime (SM)}-b_{SU(2)}^{ (SM)}\right)  \log\left[\frac{{p}^{2}}{{\mu}^{2}}\right]
& \mu,p\gg f_c\\
 b_{SU(2)}^{ \prime (SM)} \log\left[\frac{{p}^{2}}{{\mu}^{2}}\right]
 & \mu,p\ll f_c\\
 \left( b_{SU(2)}^{ \prime (SM)}-b_{SU(2)}^{ (SM)}\right)  \log\left[\frac{{p}^{2}}{{\mu}^{2}}\right]
+
 b_{SU(2)}^{ (SM)} \log\left[\frac{{p}^{2}}{{f_c}^{2}}\right]
& \mu\gg f_c\gg p.
\end{array}
\right.
\end{equation}
\end{widetext}
These limits have the following interpretation. The first $ \mu,p\gg f_c$ limit is when the theory is 
in the UV scale invariant regime. Thus the vacuum polarisation contribution to the $\beta$-function is such that
it is precisely cancelled by the SM vertex correction. The second limit corresponds to the effective theory in the IR and is just the normal 
SM vacuum polarisation contribution to the $\beta$ function. The third limit, $ \mu\gg f_c \gg p$, corresponds to working in the complete 
theory with all the heavy $f_c$ degrees of freedom still present. Again the first term is precisely cancelled by the SM vertex correction, but the 2nd term 
corresponds to the finite logarithmic correction to the $\beta$ function that one would expect to arise between the scale $f_c$ (below which scale invariance is 
first broken) and the momentum, $p$. 

Next we can insert this vacuum polarisation into the two-loop $W$ boson contribution to the Higgs mass. 
But upon performing the integral over $p$ we \emph{now} see
(by expanding in $m_{W}^{2}=g_{W}^{2}\phi^{2}$) that the Higgs mass generated
by (\ref{eq:mh2-1}) is still logarithmically divergent%
\footnote{Note the advantage that supersymmetry has: there reducing all divergences
to logarithmic ones is consistent with a solution to the hierarchy
problem. %
}. Indeed the two-loop contribution to the potential can be written
\begin{equation}
\delta V=\frac{3g^{4}}{16\pi^{2}}f_{c}^{4}\left[b_{SU(2)}J_{1}-\frac{4(N_{\tilde{\eta}}-N_{\eta})}{3}K_{1}\right]\label{eq:mh2-1-1}
\end{equation}
where \begin{widetext}
\begin{eqnarray}
J_{1} & = & \frac{1}{(4\pi)^{d/2}\Gamma(d/2)}\int_{0}^{\infty}ds\frac{s^{d/2}}{(s+\hat{m}_W^{2})}\nonumber \\
 &  & \,\,\,\,\times\left\{ 2\zeta\left(s\right)\tan^{-1}\left[\frac{1}{\zeta\left(s\right)}\right]-\log s-2\zeta\left({\hat{\mu}}^2\right)\tan^{-1}\left[\frac{1}{\zeta\left({\hat{\mu}}^2\right)}\right]+\log{\hat{\mu}}^2\right\} \nonumber \\
K_{1} & = & \frac{1}{(4\pi)^{d/2}\Gamma(d/2)}\int_{0}^{\infty}ds\frac{s^{d/2}}{(s+\hat{m}_W^{2})}\nonumber \\
 &  & \,\,\,\,\times\left\{ \frac{2}{s}\left(\zeta\left(s\right)\tan^{-1}\left[\frac{1}{\zeta\left(s\right)}\right]-1\right)-\frac{2}{\hat{\mu}^2}\left(\zeta\left({\hat{\mu}}^2\right)\tan^{-1}\left[\frac{1}{\zeta\left({\hat{\mu}}^2\right)}\right]-1\right)\right\} .
\end{eqnarray}
\end{widetext}
Expanding in $\hat{m}_{W}^{2}/s$ and using the above limits  in eq.(\ref{eq:limit-approx}),
we see that the Higgs mass, $m_W^{2}$, piece of the potential is proportional
to $\delta V\sim f_{c}^{2}m_W^{2}\log{\hat{\mu}}$. Thus the hierarchy
problem is not solved in this theory, although the divergence is rendered logarithmic thanks to the
UV scale symmetry. As described in Appendix B, the UV scale invariance
is equivalent to providing a single cancelling residue that reduces
the degree of divergence of the diagram but doesn't quite render the
integral finite. While that may seem like an improvement, generally if this
divergence is allowed to persist it means that physics is still not
truly scale invariant above $f_{c}$. There is one exception to this rule (which does not apply in this case), which is when the divergence is associated with the non-zero 
anomalous dimension of the Higgs field at the UV fixed point. Then one would expect the coefficient of logarithmically divergent Higgs mass-squared terms to 
precisely match the coefficient of logarithmically divergent kinetic terms.

Such divergences can be made harmless (as in the main body of the text) if one assumes that the 
theory runs to a fixed point with the fields having small finite anomalous dimension $\gamma$. Then one expects terms
that go like $1/\gamma$. Alternatively and more brutally one can (as in Appendix B) add additional states that remove this divergence entirely. (The former case 
is physically similar to dimensional regularisation, while the latter is similar to PV regularisation.)  In either case, there must be additional
degrees of freedom to render the Higgs mass truly scale invariant in the UV allowing one to take the continuum limit. Consequently, the hierarchy problem
remains. 

In terms of the Exact Renormalisation Group the interpretation would be as follows \cite{Rosten:2010vm}. We are envisaging a flow emanating from
a critical surface along a renormalisable trajectory. Such a flow would begin at a fixed point in which the action 
$S[\varphi,\mu,\Lambda_\infty]= S^* [\varphi]+{\cal O}(\mu/\Lambda_\infty)$, with a necessary (but not sufficient) condition for the 
theory to be renormalisable being that $\lim _{\Lambda_\infty\rightarrow\infty}(S[\varphi ,\mu,\Lambda_\infty])= S^* [\varphi] $\footnote{There is an additional 
requirement that the theory should also be self-similar, for which one should also include renormalons in the $\Lambda_\infty \rightarrow\infty$ limit (see \cite{Rosten:2010vm}).}. In order for the theory to approach the fixed point perturbatively, 
additional resonances have to remove any $\Lambda_\infty\rightarrow \infty$ divergences. Of course the usual arguments of asymptotic safety would 
then counter that maybe the theory just happens to flow close by a perturbative fixed point -- about which we are doing our perturbation theory -- and that it actually emanates from a different fixed point higher-up; the conclusion of lack of renormalisability is therefore incorrect. Unfortunately, while that may be true,  
in the present context this argument does not save us because it would imply a rapid change in the gauge beta functions around the scale of the Gaussian fixed point
where the theory is diverted to the second, higher, fixed point; one would expect this to be  
gauge mediated to the visible sector, so that the RG scale of the intermediate fixed point rather than $f_c$ would become the typical scale of relevant operators induced in 
the IR theory. 

\section{From dimensional to PV to cut-off regularisation}

For clarification (and as background to the discussion in Appendix A) 
it is useful to consider the relation between different regularisation schemes, and to ask if there is a preferred choice for scale invariant theories. As a warm-up, let us take the basic (Wick rotated)
one-loop integral
\begin{equation}
J=\int\frac{d^{4}p}{(2\pi)^{4}}\frac{1}{(p^{2}+m^{2})^{n}}.
\end{equation}
In dimensional regularisation in $d=4-\epsilon$ dimensions, and replacing
$p^{2}\rightarrow s$, this becomes 
\begin{equation}
J=\frac{\pi^{d/2}}{(2\pi)^{d}\Gamma(d/2)}\int_{0}^{\infty}ds\frac{s^{1-\epsilon/2}}{(s+m^{2})^{n+1}}.
\end{equation}

\begin{figure}
\noindent \begin{centering}
\includegraphics[scale=0.26]{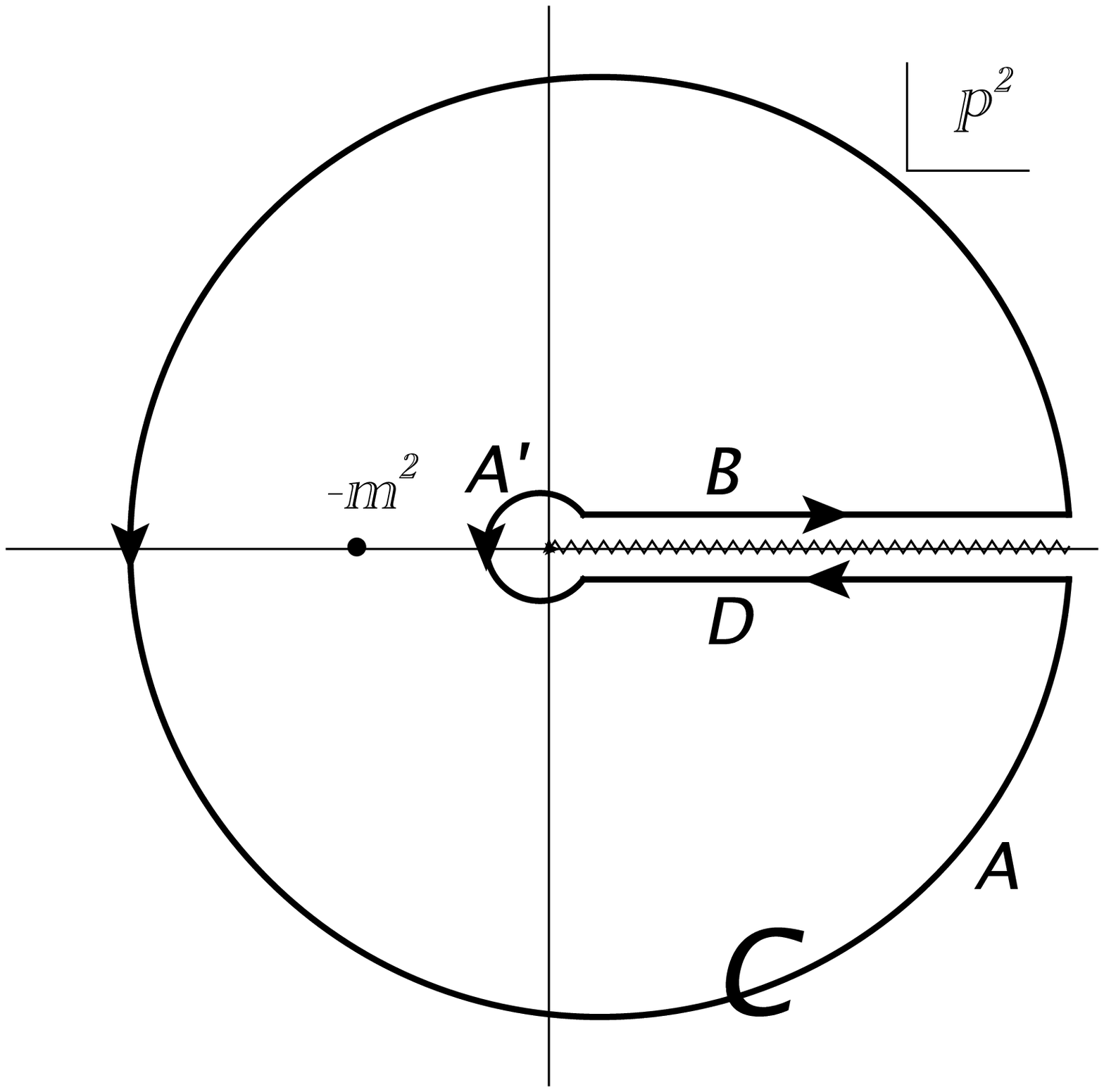}~\includegraphics[scale=0.26]{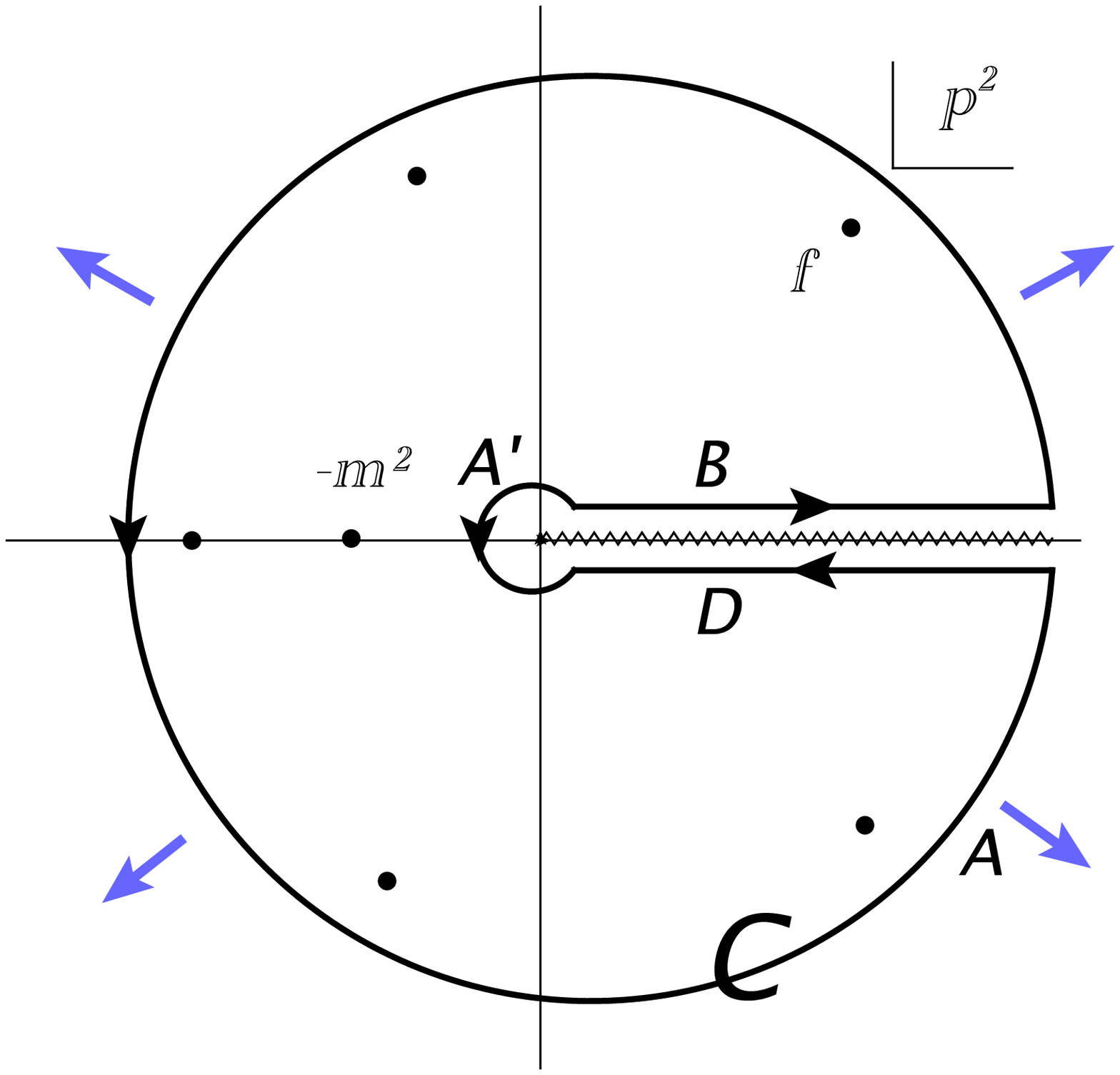}
\par\end{centering}
\caption{Keyhole contour for turning dimensional into cut-off regularisation,
left. And, right, additional cancelling singularities appearing in a
hypothetical UV complete theory allowing the arc to be taken to infinity.\label{fig:Keyhole-contour-for}}
\end{figure}

The integrand has a pole at $s=-m^{2}$, and a branch-point at
$s=0$, so one way to evaluate it is to use the ``keyhole'' contour
integral in figure \ref{fig:Keyhole-contour-for} with a branch-cut
placed along the positive real axis. Because of the branch cut, 
\begin{equation}
\int_{B\oplus D}ds\frac{s^{1-\epsilon/2}}{(s+m^{2})^{n+1}}=(1-e^{-\pi i\epsilon})\int_{0}^{s_{\infty}}ds\frac{s^{1-\epsilon/2}}{(s+m^{2})^{n+1}}.\nonumber 
\end{equation}
The $A'$ integral vanishes for any $n$, while the integral from
the arc of radius $s_{\infty}$ (denoted $A$) is of order $s_{\infty}^{1-n-\epsilon/2}$. 

The order in which limits are taken is the central issue. Suppose we
take the $s_{\infty}\rightarrow\infty$ limit first, setting the
arc integral to zero (for large enough $\epsilon$). We then have
\begin{equation}
(1-e^{-\pi i\epsilon})J=\frac{\pi^{d/2}}{(2\pi)^{d}\Gamma(d/2)}\oint_{C}ds\frac{s^{d/2-1}}{(s+m^{2})^{n+1}}.\label{eq:piJ}
\end{equation}
The integral $J$ can then be easily evaluated by the residue theorem,
and the relevant $\epsilon$-pole is already explicit in eq.(\ref{eq:piJ}).
For example when $n=1$ we have 
\begin{eqnarray}
J & = & \frac{2\pi i}{(1-e^{-\pi i\epsilon})}\frac{1}{(4\pi)^{2-\epsilon/2}\Gamma(1-\epsilon/2)}(-m^{2})^{-\epsilon/2}\nonumber \\
 & = & \frac{1}{16\pi^{2}}\left[\frac{2}{\epsilon}-\gamma_{E}-\log\left(\frac{m^{2}}{4\pi}\right)+\mathcal{O}(\epsilon)\right],
\end{eqnarray}
as usual.

So we see that $\epsilon$-poles in dimensional regularisation are
inevitable because of the branch cut. However, what about finite integrals
such as 
\begin{equation}
J=\int\frac{d^{4}p}{(2\pi)^{4}}\frac{1}{(p^{2}+m^{2})}\left(\frac{f_{c}^{2}}{p^{2}+f_{c}^{2}}\right)^{n+1},\label{eq:regint}
\end{equation}
where $n\in Z$, $n>0$ and $f_{c}\gg m$? As an example, such an integral could conceivably occur if one is considering contributions to beta functions that have 
multiple cancellations of both leading (as in Appendix A) and sub-leading terms. The integral is already
regulated by the second factor in the integrand, but the above argument
tells us that there will be $\epsilon$-poles anyway. In fact we now have contributions from two residues, one from the pole
at $s=-m^{2}$, and one from the pole at $s=-f_{c}^{2}$. In
total we find
\begin{eqnarray}
(1-e^{-\pi i\epsilon})J & = & 2\pi i\frac{\pi^{\frac{d}{2}}f_{c}^{2n}}{(2\pi)^{d}\Gamma(d/2)}\left\{ \frac{1}{n!}\partial_{s}^{n}\frac{s^{\frac{d}{2}-1}}{(s+m^{2})}|_{s=-f_{c}^{2}} \right. \nonumber \\
&& \left.\qquad\qquad  +\frac{s^{\frac{d}{2}-1}}{(s+f_{c}^{2})^{n+1}}|_{s=-m^{2}}\right\} \nonumber \\
 & = & \mathcal{O}(\epsilon).
\end{eqnarray}
The integrand is a multipole in which the residues
cancel at leading order in $\epsilon,$ leaving no $\epsilon$-pole
for $J$ but just a trail of finite pieces from the regularisation. 
This is always the case if one chooses a regulating function that does
not introduce essential singularities or branch cuts, and is similar to 
Pauli-Villars type regularisation.

Let us now return to the arc (both mathematically and metaphorically)
of the previous divergent example, and suppose that we instead expand
the propagator in $m^{2}/s$, but do not take the $s_{\infty}\rightarrow\infty$
limit before the $\epsilon\rightarrow0$ limit. For the $n=1$ integral,
we then have in total 
\begin{equation}
J=\frac{1}{16\pi^{2}}\left[\frac{2}{\epsilon}-\gamma_{E}-\log\left(\frac{m^{2}}{4\pi}\right)+\mathcal{O}(\epsilon)\right]-J_{A}
\end{equation}
where $J_{A}$ is the additional contribution from the arc, given
by
\begin{eqnarray}
J_{A} \hspace{-0.2cm}& = & \hspace{-0.2cm}\frac{2\pi i}{(1-e^{-i\pi\epsilon})}\frac{\pi^{\frac{d}{2}}}{(2\pi)^{d}\Gamma(d/2)}s_{\infty}^{\frac{d}{2}-1-n}\int_{0}^{2\pi}\frac{d\theta}{2\pi}\, e^{i(\theta(\frac{d}{2}-n-1)}\nonumber \\
 & = & \hspace{-0.2cm}\frac{1}{16\pi^{2}}\left[\frac{2}{\epsilon}+1-\gamma_{E}-\log\left(\frac{s_{\infty}}{4\pi}\right)+\mathcal{O}(\epsilon)\right].
\end{eqnarray}
This precisely cancels the pole in $\epsilon$ giving
\begin{equation}
J=\frac{1}{16\pi^{2}}\left[\log\left(\frac{\Lambda_{\infty}^{2}}{m^{2}}\right)+\mathcal{O}(1)\right],
\end{equation}
and turning dimensional into cut-off regularisation. Likewise, it
is easy to check that the familiar quadratically divergent $n=0$
integral in dimensional regularisation, 
\begin{eqnarray}
J & = & \frac{2}{\epsilon}(1+i\pi\epsilon/2)\frac{1}{(4\pi)^{2-\epsilon/2}\Gamma(2-\epsilon/2)}(-m^{2})^{1-\epsilon/2}\nonumber \\
 & = & -\frac{m^{2}}{16\pi^{2}}\left[\frac{2}{\epsilon}-\gamma_{E}+1-\log\left(\frac{m^{2}}{4\pi}\right)+\mathcal{O}(\epsilon)\right],
\end{eqnarray}
is, upon reversing the order of limits, turned into the cut-off answer,
\begin{equation}
J=\frac{\Lambda_{\infty}^{2}}{16\pi^{2}}+\frac{m^{2}}{16\pi^{2}}\log\left(\frac{m^{2}}{\Lambda_{\infty}^{2}}\right).
\end{equation}

The reason this geometric picture is useful to bear in mind is that
the residues encode the contribution to the integral from both the
low energy theory and from states of mass $f_{c}$ that we explicitly
add into the theory in order to achieve UV scale invariance, while
the arc encodes the remaining UV sensitivity (which we would obviously
like to vanish). For example, with the regulated integral example
of (\ref{eq:regint}) the latter contribution vanishes as the radius
is taken to infinity, as in fig.\ref{fig:Keyhole-contour-for}. 

Suppose now that we have a theory that emanates from a fixed point in the
UV, so we decide to measure all our couplings at some high scale $\mu\gg f_{c}$
above which we consider $g$ to be roughly constant. Then in the sliding-scale
scheme discussed in Appendix A we would write an expression like eq.(\ref{eq:master}) with
${C}_{tot}$ (or more precisely the $\beta$-function) vanishing at $\hat{p}=\hat{\mu}$. 
The difference between the result at $\mu$ and the would-be continuum result is then given by subtracting 
one contour integral from the other which leaves a 
contour integral round an annulus (with branch-cut) with inner radius $s_\mu=\hat{\mu}^2$ and outer radius $s_\infty = \hat{\Lambda}^2_\infty$. 
In order to be able to take the continuum limit one would require this integral to converge to a constant 
as  $\Lambda_\infty \rightarrow \infty$ with $\mu \gtrsim f_c$. 
Conversely, if the arc integral blows up in the $\Lambda_{\infty}\rightarrow\infty$
limit, a continuum limit does not exist within the perturbative description being considered: despite the theory becoming scale invariant
in the UV, the higher the scale at which one measures the couplings,
the more precisely one has to do it in order to have control over
the low energy theory. From the point of view of the exact renormalisation group, one would say that 
the theory never reaches the perturbative (but non-trivial) fixed point we had in mind, but flows past it, perhaps to some other interacting 
fixed point. 

In addition we conclude that dimensional regularisation
has no preferred status in theories based on exact UV scale invariance. Indeed, suppose that one actually \emph{had} a UV completion
in which all integrals were cancelled by resonances and the arc really
could be taken to infinity. In such a theory the $s_{\infty}\rightarrow\infty$
and $\epsilon\rightarrow0$ limits would have to commute since regularisation
is unnecessary. The only conceivable advantage of dimensional
regularisation would then be if, in conjunction with the Coleman-Weinberg prescription
of setting tree-level mass-terms to zero, what was left after the
arc is taken to infinity were somehow a better approximation to the
IR physics than simply putting a cut-off at $f_{c}$. As is evident
from the main body of the text, this is generally not the case, so we conclude that
``classical scale invariance'' is not a good guiding principle for discussing the phenomenology of spontaneously broken exact scale invariance. 
All of which is not to say that the Coleman-Weinberg mechanism has no place, but just that it is a different program from that of exact UV scale invariance.

\end{appendices}

\end{document}